\documentclass[preprint,preprintnumbers,floatfix,12pt]{revtex4-1} 
\usepackage{times}
\usepackage{epsfig}
\usepackage{geometry}                		
\usepackage{graphicx}
\usepackage{xcolor}
\usepackage{float}
\usepackage{amsmath,amssymb}
\usepackage[titletoc,toc]{appendix}	
\usepackage{subfig} 
\usepackage{bm}
\usepackage[bottom]{footmisc}

\begin{document}

	\title{Electrolytes in regimes of strong confinements: surface charge modulations, osmotic equilibrium and electroneutrality}
	\author{Amin Bakhshandeh}
	\email{amin.bakhshandeh@ufrgs.br}
	\affiliation{Instituto de F\'isica, Universidade Federal do Rio Grande do Sul, 91501-970, Porto Alegre, RS, Brazil}
	\affiliation{Departamento de F\'isico-Qu\'imica, Instituto de Qu\'imica, Universidade Federal do Rio Grande do Sul, 91501-970, Porto Alegre, RS, Brazil}
	\author{Maximiliano Segala}	
	\email{maximiliano.segala@ufrgs.br}
	\affiliation{Departamento de F\'isico-Qu\'imica, Instituto de Qu\'imica, Universidade Federal do Rio Grande do Sul, 91501-970, Porto Alegre, RS, Brazil}
	\author{Thiago Colla}
	\email{colla@ufop.edu.br}
	\affiliation{Instituto de F\'isica, Universidade Federal de Ouro Preto, 35400-000, Ouro Preto, MG, Brazil}

	%\keywords{interfaces, induced interactions, electroneutrality, surface charge modulations}

	\begin{abstract}
		In the present work, we study an electrolyte solution confined between planar surfaces with nonopatterned charged domains, which has been connected to a bulk ionic reservoir. The system is investigated through an improved Monte Carlo (MC) simulation method, suitable for simulation of electrolytes in the presence of modulated surface charge distributions. We also employ a linear approach in the spirit of the classical Debye-H\"uckel approximation, which allows one to obtain explicit expressions for the averaged potentials, ionic profiles, effective surface interactions and the net ionic charge confined between the walls. Emphasis is placed in the limit of strongly confined electrolytes, in which case local electroneutrality in the inter-surface space might not be fulfilled. In order to access the effects of such lack of local charge neutrality on the ionic-induced interactions between surfaces with modulated charge domains, we consider two distinct model systems for the confined electrolyte: one in which a salt reservoir is explicitly taken into account {\it via} the osmotic equilibrium with an electrolyte of fixed bulk concentration, and a second one in which the equilibrium with a charge neutral ionic reservoir is implicitly considered. While in the former case the osmotic ionic exchange might lead to non-vanishing net charges, in the latter model charge neutrality is enforced through the appearance of an implicit Donnan potential across the charged interfaces. A strong dependence of the ionic-induced surface interactions in the employed model system is observed at all particle separations. These findings strongly suggest that due care is to be taken while choosing among different scenarios to describe the ionic exchanging in electrolytes confined between charged surfaces, even in cases when the monopole (non zero net charge) surface contributions are absent.  
	\end{abstract}
	
	\maketitle

	\section{Introduction}
	
	Interfaces are a topic of great relevance in a number of different research areas, ranging from physics and chemistry to biology and, more recently, nonotechnology and nano-engineering. Many reactions or physical properties can be induced by the presence of an interface, and are dependent not only on their local environments, but also on the kind of surfaces they are made of~\cite{levin,sprycha,grahame}. These surfaces might be semi-permeable, thereby controlling the flux of different components over the interface, or can also have specific interactions with different components from their vicinity, which will be either adsorbed onto or repelled from the surface. Obviously, these properties can significantly change the behavior of the system with respect to its bulk state, and might give rise to a number of interesting phenomena, with many practical applications in different areas~\cite{Israelachvili,Lozada92}. 
	
	One common example are interfaces comprised by surfaces immersed in aqueous solutions, as is the case in many biological systems~\cite{levin}. These surfaces may acquire a net electric charge because of ionization of certain a acidic or basic groups, or {\it via} the adsorption of charged molecules onto the surfaces~\cite{levin,amin2018,dos2016_1,amin2011}. The charged surface attracts oppositely dissolved ions from the environment, leading to a electrostatic screening of the surface charge through the formation of a complex charged structure generally known as Electric Double Layer (EDL). Helmholtz, back in the 1850s, was the first to study the structure of EDL, describing it as a cloud of surrounding counterions which renders the surface potential an exponential decay~\cite{Helmholtz}. A few years later, Gouy, Chapman, and Stern (GCS) also attempted to describe the properties of EDL~\cite{gouy,chapman,stern,burt2014}. In Stern's approach, the double layer was considered as a thin layer of counterions electrolytically bounded to the charged surface – the so-called Stern layer – which strongly screens the surface charge. Gouy and Chapman then addressed the presence of a further diffuse ionic layer in which ions display smooth distributions in response to the field provided by the compact Stern layer~\cite{gouy,chapman,stern}. Despite its simplicity, the GCS theory gives us a clear physical picture, still able to provide valuable insights into mechanisms underlying various interesting phenomena~\cite{levin,Israelachvili,oldham}. Later on, Derjaguin, Landau, Verwey, and Overbeek (DLVO) proposed a very powerful theory based on this simple picture, in which the interaction between EDLs is described by a combination of screened electrostatics and van der Waals forces~\cite{levin,Israelachvili,derjaguin}. 
	
	The DLVO theory has been applied to successfully describe a number of properties of interacting EDLs, including the  stability of suspended nanoparticles against irreversible aggregation~\cite{derjaguin,verwey1,marshall,missana,Bel00}, which depends on a fine balance between short range attractive van der Waals forces and screened electrostatic interactions~\cite{Ohshima,Israelachvili,ohshima2011,verwey}. One of the main assumptions behind the classical DLVO theory is that the region between the EDLs is fully balanced by oppositely charged counterions, in such a way as to keep electroneutrality in the inter-surface region. Although the long-range nature of the Coulomb potential requires {\it overall} charge neutrality in three dimensions to be fulfilled, the assumption of {\it local} charge neutrality in the inter-surface region should not be enforce {\it a priori}, as the mobile neutralizing counterions are free to diffuse throughout the system. The implications of this lack in local charge neutrality on induced surface interactions have been first addressed by Lozada-Cassou and co-workers~\cite{Cas96,Loz96,Agui02,Agui02}. This point has recently attracted renewed attentions~\cite{colla,Levy20} thanks to experimental work of Luo {\it at. al.}, which demonstrated the absence of charge neutrality in the region between charged colloidal surfaces~\cite{Luo}. The electroneutrality condition is generally accepted as a natural assumption when the confined system has no contact with its external surroundings, as the charged surfaces release their own neutralizing counterions into the confining region. However, the situation changes if the nano-confined electrolyte is allowed to exchange ions with its external environment -- as might be the case in nano-pores, nono-sizes membranes or narrow connecting channels -- in which case the local charge neutrality might be  interpreted as a simplifying assumption rather than a necessary condition. In this situation, the external ionic reservoir has to be always able to supply the closed system with the necessary amount of counterions to neutralize the charged surfaces, regardless of the inter-surface space. Since the reservoir itself must have zero net charge and sustain no field, this has to be accomplished at the cost of the building-up of a potential difference between the confined system and the ionic reservoir across their interface -- the so-called {\it Donnan potential}~\cite{Don24}. The resulting osmotic equilibrium is named the {\it Donnan equilibrium}, and follows from the usual chemical equilibrium between the ionic spices, in addition to the electroneutrality condition in both confined and external systems~\cite{Ohshi85,Tam98,Jim04,Bryk06,Wang09}. Notice that this approach completely neglects ionic correlations across the interfaces, which are known to be present in real systems~\cite{Loz96_2,Loz97,Deg98}.
	
	Another limitation of the classical DLVO theory relies on the fact that it explicitly assumes uniform charge distributions all over the charged surfaces. Even though the inclusion of such effects might significantly increase the complexity in the system description, realistic approaches of the interactions of charged surfaces requires the incorporation of such effects, as many of the technologically relevant charged systems are comprised of patchy-like surface charge domains. One example are charged surfaces made of nano-patterned charge modulations~\cite{Bakhshandeh2019}, which can be designed {\it via} nano-fabrication techniques~\cite{part2005,sayin}. These systems have recently attracted a lot of attention due to their potential application in the production of nano-technological devices and biological systems~\cite{teshome2014}. On the other hand, inclusion of such multipole contributions to the surface charge significantly increases the numerical complexity in the description of EDLs, as the symmetry breakdown across the surface parallel directions prevents the usage of powerful simplifying approaches such as the Gauss' Law. Despite these drawbacks, many efforts have been made over the past decades to incorporate effects of charge inhomogeneities on the classical CGS model, and considerable improvements over the traditional approaches have been achieved ~\cite{Mik94,Mik95,Whi02,Ben07,Sil12,Mad13,Ben13,aminbkh,Gho17,Adar18,aminb2018,Sam19,Zhou20}.  
	
	Simulation of electrolytes in contact of patterned charge modulations is also a rather challenging task, because point charged particles should be included on the surfaces in such a way as to construct the target non-uniform configurations~\cite{aminbkh,DoGi16}. Since the whole system must be periodically replicated, this leads to an infinite summation over the replicas, which normally is  performed using Ewald techniques~\cite{toukma1}. The drawback of this method is that, in order to have continuous charge distribution on the surface, one should include a very large number of point charges on the plate which, in turn, considerably slows down the simulation's time. Considerable gain in simulation efficiency can be achieved by employing advanced simulation techniques that circumvent this problem, such as the replacement of the inhomogeneous charge by a discrete array of point charges placed in a suitable location behind the surfaces \cite{Mor02}. Recently, an alternative method has been proposed which also allows us to simulate non-uniformly charged surfaces using much less CPU time~\cite{aminb2018}. The key idea in this approach is to treat the continuous electrostatic potential produced by a periodic surface charge distribution separately from the one produced by the mobile ions. The implementation of this method requires that the overall inhomogeneous charge on the plate should be zero~\cite{aminb2018}. The method can therefore be applied in cases of patterned-like charge distributions with periodic charged domains, which keeps the non-homogeneous surface charge distribution globally neutral.
	
	In the present work, we apply the aforementioned MC method in combination with the method proposed in Ref.~\cite{DoGi16} for studying electrolytes confined by charged surfaces comprised of periodic charged domains in addition to a uniform charge background. The system is also investigated in the framework of a linear, Debye-H\"uckel (DH) approximation, which has the advantage of providing analytical expressions that give insights into the role of different parameters on the main physical mechanisms. In order to investigate the interplay between charge neutrality and effective interactions between the surfaces, we consider the context of two distinct model systems for the confining electrolyte: one in which a free particle exchange with an ionic reservoir is explicitly allowed, and a second one in which charge neutrality is imposed on both the confined electrolyte and the external reservoir (the so-called Donnan approach). We show that the implementation of these models to similar physical systems leads might to quite different behaviors for the surface forces, even in the context of a linear approximation which neglects various key contributions to these systems. 
	
	The paper is organized as follows. In section~\ref{model}, the model systems applied to describe an electrolyte confined between charged surfaces is described. Next, in Section \ref{sim}, the simulation details are discussed, and a method is described that enables us to simulate these systems more efficiently. In section~\ref{theory}, we apply a linear approximation to describe the general system properties in the context of the two proposed model systems. Results are then presented and discussed in some detail in Sec.~\ref{results}. Finally, conclusion remarks and perspectives are outlined in ~\ref{conclusions}, followed by the Appendix, where technical aspects regarding force calculations and averaged potential are worked out in detail.

	\section{MODEL SYSTEM}\label{model}
	
	We consider an electrolyte confined in the region between two flat, charged surfaces. We adopt a coordinate system with origin in the middle point between the surfaces. The surfaces $1$ and $2$ are located at positions $z_1=-d/2$ and $z_2=d/2$, respectively, and possess inhomogeneous surface charge distributions $\sigma_1(x,y)$ and $\sigma_2(x,y)$, respectively ($x$ and $y$ are in-plane coordinates). The surfaces are infinitely thin, but are covered by membranes of width $z_m$ on its both sides, which avoids the penetration of ions and sets in the ion-surface closest contact. 
	
\begin{figure}[h!]
		\includegraphics[width=8.75cm,height=3.75cm]{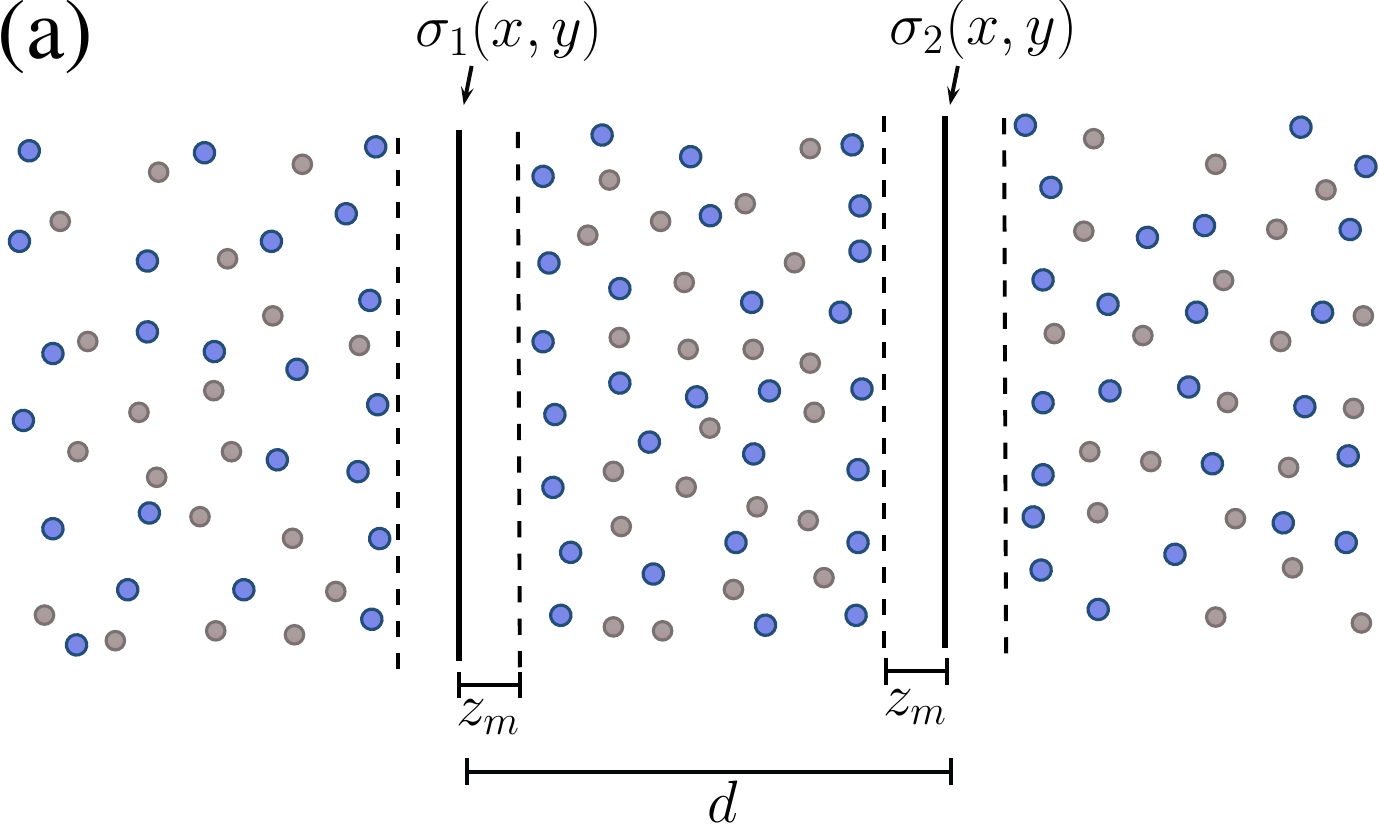}\\
		\includegraphics[width=8.75cm,height=3.75cm]{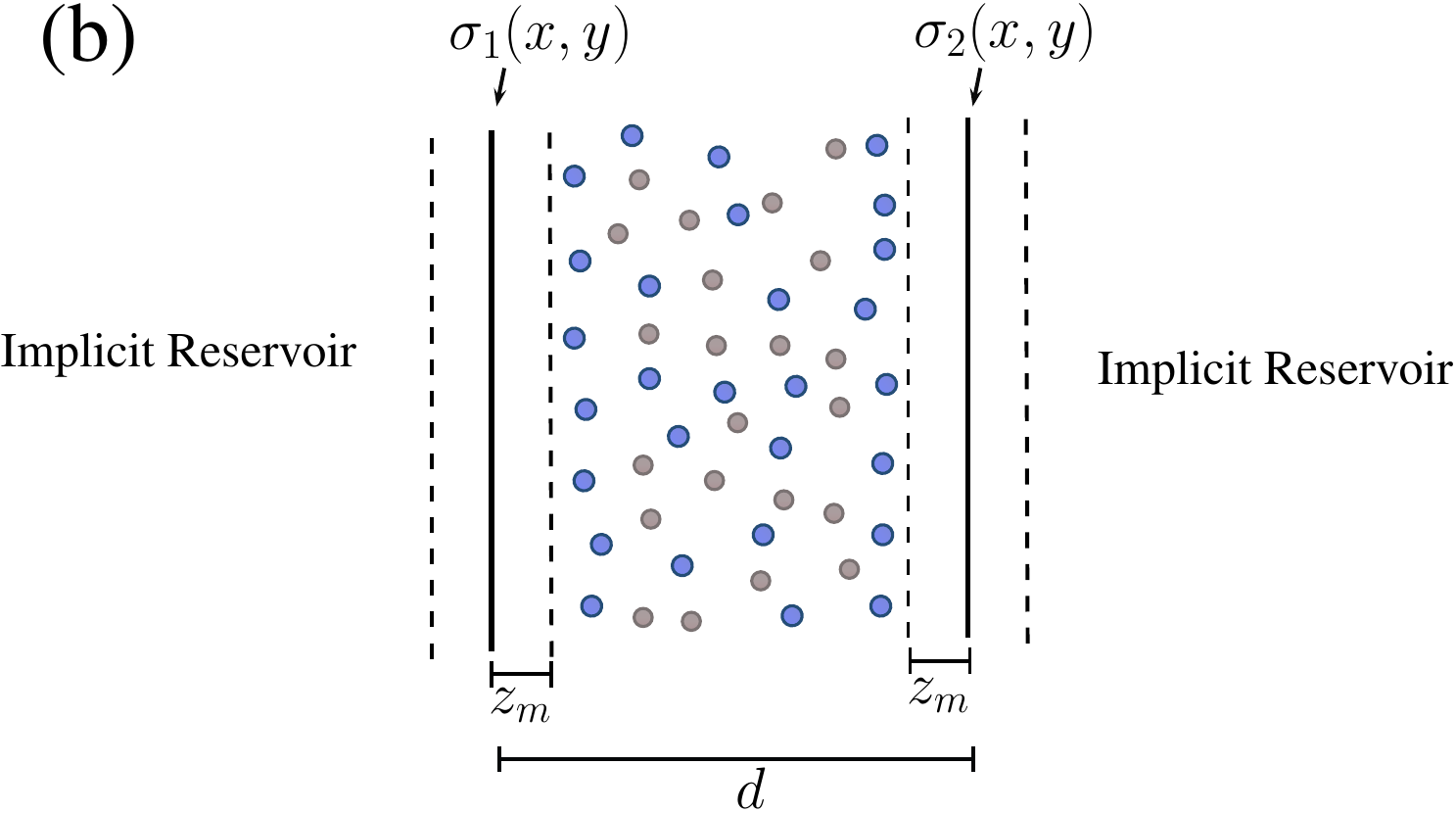}
		\caption{Model systems applied to describe an electrolyte confined between charged surfaces of charges $\sigma_1(x,y)$ and $\sigma_2(x,y)$, separated by a distance $d$ and covered by a membrane of thickness $z_m$. In the case (a), a free ionic exchange between an explicit ionic reservoir of mean concentration $c_s$ located at the external region is considered (explicit reservoir model). In the second situation (b), the ionic reservoir is not considered explicitly, and the confined net charge is zero.}
		\label{fig:fig1}
	\end{figure}

	Two distinct scenarios will be considered for modeling the confined electrolyte. In the first case (implicit reservoir), we shall consider the confined electrolyte to be in osmotic equilibrium with an electrolyte of given bulk concentration $c_s$, which is {\it explicitly} located away from the plates. Ionic diffusion is freely allowed through the membranes, although ionic penetration into the membrane region is avoided. This model system is sketched in Fig.~\ref{fig:fig1}a. 
	
	In the second model, the system is in equilibrium with an {\it implicit} charge reservoir of zero local charge. In this so-called {\it Donnan model}, ion flux is allowed, but the inter-surface region should be always charge-neutral, as well as the region just beyond the surfaces. This is accomplished {\it via} the emergence of a potential difference between the system is its implicit reservoir. This model system is depicted in Fig.~\ref{fig:fig1}b.
	
	\section{MONTE CARLO SIMULATIONS}\label{sim}

\begin{figure}[h!]
		\begin{center}
			\includegraphics[width=9cm,height=4.5cm]{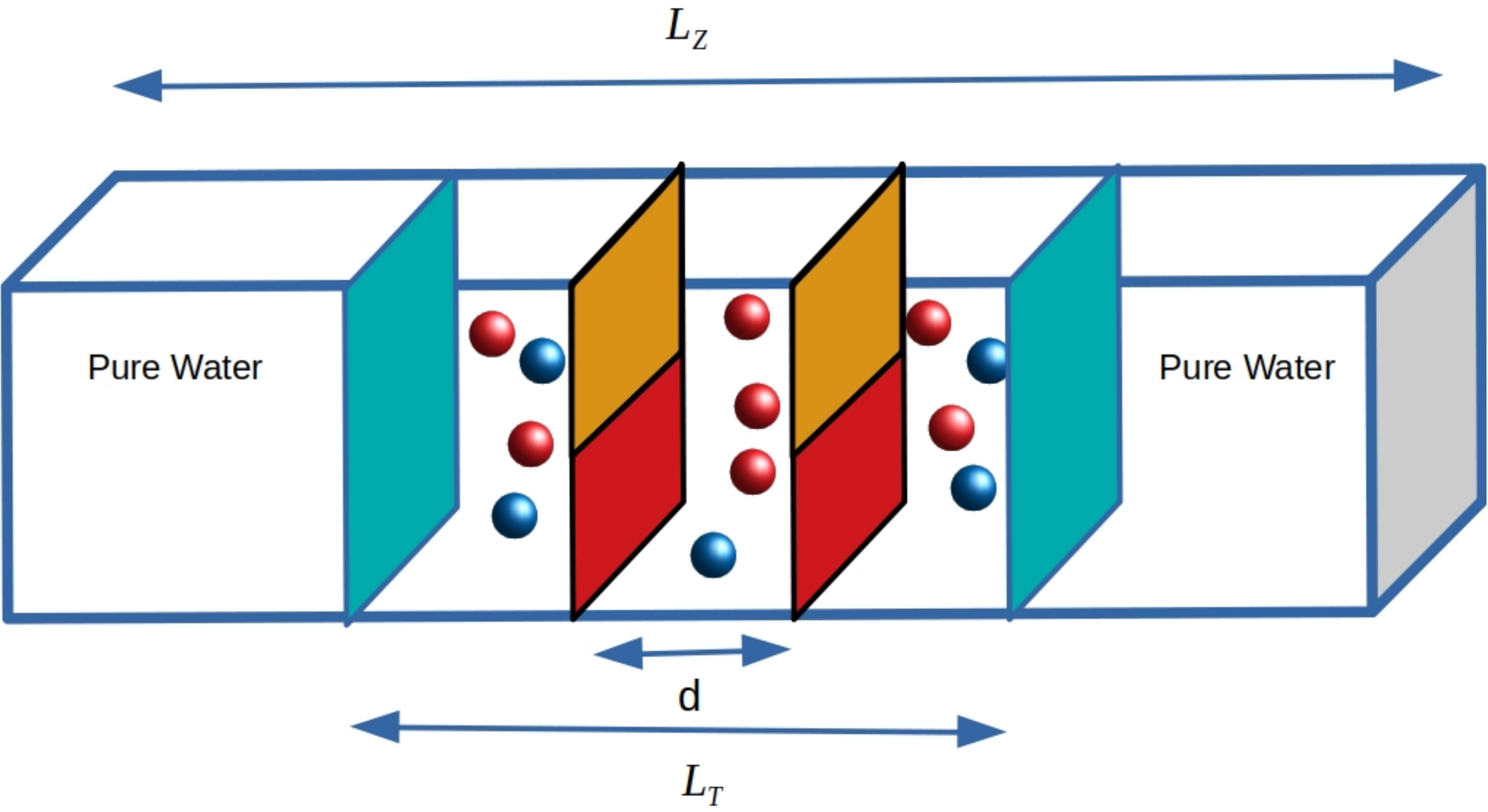}
		\end{center}
		\caption{Representation of the cell used in our simulations. The longitudinal $z$-axis is perpendicular to the patterned charged surfaces and has its origin at the mid-plane. The plates are located at $z_1 = -d/2$ and $z_2 = d/2$, while the $x-y$ plane lies parallel to the surfaces. Ions are located at the region $|z|<L_T$, and are free to move between the inter-surface space $|z|<d/2$ and the bulk phase at $d/2<|z|<L_T/2$. The regions $L_T/2<|z|<L_z/2$ are devoided from ions, and contain only the continuum solvent. This region has no direct correspondence to our physical system, and is present only to avoid cell replications along the $z$ direction.} 
		\label{fig:fig2}
	\end{figure}

	We consider an electrolyte confined in our simulation box with its natural parameters, the same as in Ref.~\cite{colla}, as is depicted in Fig.~\ref{fig:fig2}. The electrolyte solution is placed in the region $ -L/2 \le x \le L/2$, $-L/2 \le y \le L/2$, where $x$ and $y$ are the in-plane, transversal coordinates. The plates are located at $z_1=-d/2$ and $z_2=d/2$, where $d$ is separation distance between plates, and $z$ is the coordinate orthogonal to the flat walls. The width and length of the plates are set to be  $L_x = L_y = L = 5$~nm, respectively.
	The total size of the cell across the orthogonal direction is $L_z = 25$~nm. The regions $-L_z /2 \le z \le -L_T /2$ and $L_T /2 \le z \le L_z /2$, contain pure solvent (water), where $L_T =L_z/2$. In the present study, water is modeled as a continuum of dielectric constant $\varepsilon_w = 80 \varepsilon_0$ ($\varepsilon_0$ is the vacuum permitivity) and the temperature is fixed at $T=300$ K, while the ions are modeled as hard spheres of hydrated radius $2$~\AA. It is important to point out that the solvent compartments at the regions $|z|>L_T/2$ have no physical relevance in the context of the model system described in the previous section. The inclusion of these regions here is just an {\it artifact} of the simulation technique employed, and aims to avoid replication of the simulation box across the $z$ axis, orthogonal to the wall's plane.
	
	The simulation of bulk electrolytes in the presence of flat surfaces with inhomogeneous charge distributions is difficult due to the unavoidable presence of neighboring cells. One way to avoid the time consuming summation over a large number of surface charges is to consider the effects from an array of point charges placed {\it behind} the charged surfaces, as is done in Ref. \cite{Mor02}. In this method, point charges are put at a specific distance behind the plates, and periodic boundary conditions are then imposed on the system by employing the Lekner-Sperb's method ~\cite{lekner1991,sperb1998}. Here we shall consider an alternative approach which consists of replacing the field of many point-like ions by a continuous surface field, as outlined in what follows.

	For each plate we consider a simple combination of  homogeneous and periodic (sinusoidal) charge distributions, as follows~\cite{aminb2018}:
	%%%%%%%%%%%%%%%%%%%%%%%%%%%%%%%%%%%%%%%
	\begin{equation}
	\sigma(x,y) = \sigma_{0}  \left[1+\sin(K_x x+\varphi_x)  \sin(K_y y + \varphi_y)\right],
	\label{eq1}
	\end{equation}
	%%%%%%%%%%%%%%%%%%%%
	where $\sigma_{0}$ is the surface net charge density, $\varphi_{x}=0$ and  $\varphi_{y}=\pi/2$  are the phase constants across $x$ and $y$ directions, respectively, $K_x \equiv 2 \pi n_x/L_x$  , $K_x \equiv 2 \pi n_y /L_y$ are the periodic wavenumbers,  $n_{x}$ and $n_{y}$ being integers that denote the modulation sites over the $x$ and $y$ directions, respectively. 
	The electrostatic potentials due to the sinusoidal surface charge modulations in Eq. (\ref{eq1}) can be written as:~\cite{aminb2018}
	
	%%%%%%%%%%%%%%%%%%%% 
	\begin{eqnarray}
	\Phi(\bm{r} ) & = &  \dfrac{2 \pi \sigma_0}{\varepsilon_w K} \sin(K_x x + \varphi_x)  \cos(K_y y + \varphi_y)   e^{-K\lvert z \rvert},
	\label{eq2}
	\end{eqnarray}
	%%%%%%%%%%%%%%%%%%%%
	where $K\equiv\sqrt{K_x^2+K_y^2}$. The corresponding multipole contribution $\Phi_m(\bm{r})$ from the two flat surfaces represented in Fig.1, $\Phi_m(\bm{r})$ is thus the superposition
	\begin{eqnarray}
	\Phi_m(\bm{r} )=  \frac{2 \pi \sigma_0}{\varepsilon_w K} \sin(K_x x +\varphi_x)  \sin(K_y y + \varphi_y)\left( e^{-K \lvert z+z_1/2 \rvert} + e^{-K\lvert z-z_2/2 \rvert} \right).
	\label{eq3}
	\end{eqnarray}
	The total electrostatic potential produced by the parallel surfaces is the simple superposition of their multipole and the uniform charge distributions: 
	\begin{equation}
	\Phi_{t}(\bm{r} )=\Phi_m(\bm{r} )+\Phi_h(\bm{r} ),
	\label{eq4}
	\end{equation}
	%%%%%%%%%%%%%%%%%%%%  
	where  $\Phi_h(\bm{r})$ is potential due to
	homogeneous charge distribution surfaces, which is given by~\cite{colla}:

	\begin{equation}
	\Phi_h(\bold{r} )=\left\{
	\begin{array}{@{}ll@{}@{}}
	\dfrac{4 \pi}{\varepsilon_w} \sigma_0 (z+z_1/2)  & z<-z_1/2, \\
	0& -z_1/2<z<z_2/2,\\
	\dfrac{4 \pi}{\varepsilon_w} \sigma_0 (z-z_2/2) & z>z_2/2,
	\end{array}\right.
	\label{eq5}
	\end{equation}

	The total system energy comprises ion-ion as well as ion-surfaces interactions, and can be written as:~\cite{colla,DoGi16}
	\begin{eqnarray}\label{ener}
	U=\sum_{{\pmb k}\neq{\pmb 0}}^{\infty}\frac{2\pi}{\varepsilon_w V |{\pmb k}|^2}\exp{\left(-\dfrac{|{\pmb k}|^2}{4\kappa_e^2}\right)}[A({\pmb k})^2+B({\pmb k} )^2] + \nonumber \\
	\frac{2\pi}{\varepsilon_w V}\left(M_z^2 -Q_t G_z \right) + \dfrac{1}{2}\sum_{i \ne j}^Nq_iq_j\frac{\text{erfc}(\kappa_e|{\pmb r}_i-{\pmb r}_j|)}{\varepsilon_w |{\pmb r}_i-
		{\pmb r}_j|}+\nonumber \\
	\sum_{i=1}^{N}q_i\Phi_t ({\pmb r}_i ) \ ,
	\label{eq6}
	\end{eqnarray}
	where the coefficients just introduced above are given by:
	\begin{subequations}
		\begin{align}
		A({\pmb k})&=\sum_{i=1}^N q_i\text{cos}({\pmb k}\cdot{\pmb r}_i) \ ,  \\
		B({\pmb k})&=-\sum_{i=1}^N q_i\text{sin}({\pmb k}\cdot{\pmb r}_i) \ ,  \\
		M_z&=\sum_{i=1}^N q_i z_i \ ,  \\
		G_z&=\sum_{i=1}^N q_i z_i^2 \ ,  \\
		Q_t&=\sum_{i=1}^N q_i,
		\label{eq7}
		\end{align}
	\end{subequations}
	where $V$ is the volume $ L_x \times L_y \times L_z$, which includes the vacuum region of the modified Ewald method, $k_e$ is the damping parameter (here taken to be $ 4/L_x$), and the $k$-vectors are $k = (2\pi w_x/L_x,\\ 2\pi w_y/L_y, 2\pi w_z /L_z)$,
	where $w$’s are integers.

	We should mention that in Eq.~\ref{eq6} the effect of other plates in the neighboring cells has not been considered. At this stage, we should check whether other plates in neighboring cells can perturb the distribution of the ions inside the main cell. To address this issue, we implemented MC simulation for plates where $n_x$ and $n_y$ are $1$ and $0$, respectively, considering two distinct situations. In one case, the charge distributions are discontinuous, and are obtained by distributing a big number of point charges on the surfaces in such a way that the charge distribution in Eq. \ref{eq1} is satisfied. The charge of particles is then tuned to obtain the overall $\sigma_0$ for each plate and the energy is evaluated by using the modified Ewald summation method. As a result, the energy contains the effect of other plates in other neighboring cells. In the other case (of continuous charge distributions) we explicitly consider the analytical surface potentials as described above. The plates are placed in the presence of an electrolyte containing mono or multivalent ions, such that their density profiles can be computed after equilibration is achieved. To reduce the time of simulations we put only $400$ point charges on each plate.
	
\begin{figure}[h!]
		\begin{center}
			\includegraphics[width=8cm]{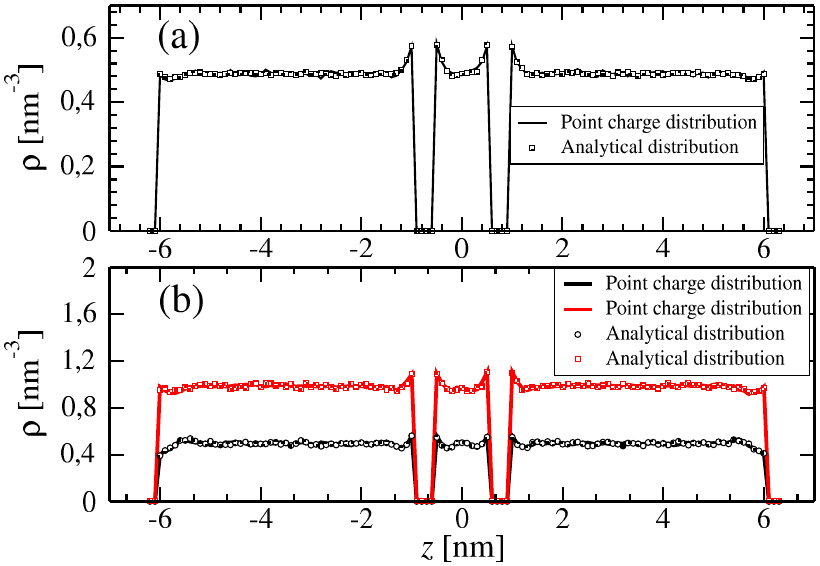}
		\end{center}
		\caption{Ionic profiles for ions in the presence of surfaces with point charges and continuous charge distribution. The number of particles on the surface for the case of discontinuous distributions is $400$. $\sigma_0 = 0.1 $ C/m$^2$ for two cases. Solid lines are ion's distribution in presence of discontinues charge distribution and symbols are density profile for continuous charge distribution. (a) Density profile for mono valence salt, (b) density profile for $2:1$ salt.}  
		\label{fig:fig0}
	\end{figure}

	Simulations performed using a Canonical Monte Carlo (CMC) algorithm~\cite{allen,Frenkel,metro}.  Equilibration is achieved with $10^6$ MC steps and each $10^5$ uncorrelated particle configuration is saved for analysis. Since our system is in contact with an ionic reservoir of fixed concentration $c_s$, we perform simulations in an iterative fashion, in which the number of mobile ions in the simulation box is adjusted until the equilibrium ionic profiles reach their bulk value $c_s$ characteristic of the equilibrium with a salt reservoir of same concentration. In Fig.\ref{fig:fig0},  we compare the density profiles resulting from continuous (solid lines) and discontinuous (symbols) surface charge distributions, for a surface charge of  $\sigma_0=0.1$ C/m$^2$ in the presence of both monovalent (a) and multivalent (b) electrolytes. An excellent agreement is observed between the two approaches.
	
	%%%%%%%%%%%%%%%%%%%%%%%

	%Nevertheless, to see the effect of background charge on the patterned surface, we put another $400$ point charge randomly on each surface with an equal charge to obtain an additional $\sigma_0$ on each plate, in this case, the plates are not neutral anymore. We can compare the results with complete potential as in Eq.~\ref{eq2} 
	%The results are shown in Fig.\ref{add1}, it is observed that there is a good agreement between discontinues and analytical solution is also observed in this case.  
	
	%  \begin{figure}
	%	\begin{center}
	%		\includegraphics[width=7cm]{zf.eps}
	%	\end{center}
	%	\caption{Density profiles for ions in the presence of surfaces with point charges and continuous charge distribution both with background charge. The number of particles on the surface for the case of discontinuous distributions is $400$ and for background, the charge is $400$. $\sigma_0 = 0.1$ for two cases. Solid lines are ion's distribution in presence of discontinues charge distribution and symbols are density profile for continuous charge distribution. a) Density profile for mono valence salt.}  %
	%	\label{add1}
	%\end{figure} 

	\section{THEORETICAL DESCRIPTION}\label{theory}
	
	We are now going to outline the theoretical approach applied to describe the model systems described in Section \ref{model}. As we shall shortly see, the basic properties of the confined electrolyte can be readily evaluated once the averaged electrostatic potential $\Psi(\bm{S},z)$ is calculated, where $\bm{S}=x\hat{\bm{e}}_x+y\hat{\bm{e}}_y$ is the in-plane position vector. The dimensionless averaged potential $\psi(\bm{S},z)\equiv\beta q \Psi(\bm{S},z)$ (with $\beta=(k_BT)^{-1}$ being the inverse thermal energy, and $q$ the elementary charge) has to satisfy the following Poisson equation:
	\begin{equation}
	\nabla^2\psi(\bm {S},z)=-4\pi\lambda_B\left[\sum_i\alpha_i\rho_i(z,\bm{S})+\varrho_1(\bm{S},z)+\varrho_2(\bm{S},z)\right],
	\label{eq8}
	\end{equation}
	where $\lambda_B\equiv\beta q^2/\varepsilon_w$ is the Bjerrum length, and $\alpha_i=\pm1$ is the ionic valence of ions of specie $i$. Here, $\varrho_1(\bm{S},z)\equiv \sigma_1(\bm{S})\delta(z+d/2)$ and $\varrho_2(\bm{S},z)\equiv \sigma_2(\bm{S})\delta(z-d/2)$ represent the fixed charge densities of surfaces $1$ and $2$ located at positions $z_1=-d/2$ and $z_2=d/2$, respectively, with $\sigma_1(\bm{S})$ and $\sigma_2(\bm{S})$ denoting the corresponding inhomogeneous surface charge densities placed on these surfaces. Alternatively, the presence of the walls can be incorporated into the Poisson equation through the following boundary conditions across the walls positions:
	\begin{subequations}
		\begin{align}
		&\dfrac{\partial \psi(\bm{S},z)}{\partial z}\biggr\arrowvert_{z=(d/2)^+}-\dfrac{\partial \psi(\bm{S},z)}{\partial z}\biggr\arrowvert_{z=(d/2)^-} = -4\pi\lambda_B\sigma_1(\bm{S)},&\\
		&\dfrac{\partial \psi(\bm{S},z)}{\partial z}\biggr\arrowvert_{z=(-d/2)^+}-\dfrac{\partial \psi(\bm{S},z)}{\partial z}\biggr\arrowvert_{z=(-d/2)^-} = -4\pi\lambda_B\sigma_2(\bm{S}),&
		\end{align}
		\label{eq9}
	\end{subequations}
	where $(d/2)^{\pm}$ stands for the limits $\lim _{\epsilon\rightarrow 0} d/2\pm \epsilon$. Due to the symmetry in the orthogonal direction $z$, it is convenient to consider the two-dimensional Fourier transform of the electrostatic potential along the in-plane surface coordinates:
	\begin{equation}
	\hat{\psi}(\bm{q},z)=\dfrac{1}{(2\pi)^2}\int \psi(\bm{S},z)e^{-i\bm{q}\cdot\bm{S}}d\bm{S},
	\label{eq10}
	\end{equation}
	where $\bm{q}=q_x\bm{\hat{e}}_x+q_y\bm{\hat{e}}_y$ is the in-plane wave number vector. The potential $\psi(\bm{S},z)$ can be obtained from its Fourier components $\hat{\psi}(\bm{q},z)$ {\it via} the inverse transform:
	\begin{equation}
	\psi(\bm{S},z)=\int\hat{\psi}(\bm{q},z)e^{i\bm{q}\cdot\bm{S}}d\bm{q}.
	\label{eq11}
	\end{equation}
	Similar expressions hold for the Fourier transformed surface charges $\hat{\sigma}_1(\bm{q})$ and $\hat{\sigma}_2(\bm{q})$. Substitution of Eq. (\ref{eq11}) into Eq. (\ref{eq8}) results in the following Poisson Equation for the Fourier components of the electrostatic potential:
	\begin{equation}
	\dfrac{\partial^2\hat{\psi}(\bm{q},z)}{\partial z^2}-q^2\hat{\psi}(\bm{q},z)=-4\pi\lambda_B\sum_i\alpha_i\hat{\rho}_i(\bm{q},z),
	\label{eq12}
	\end{equation}
	where $\hat{\rho}_i(\bm{q},z)$ are the Fourier components of the local ionic densities $\rho_{i}(\bm{S},z)$. Similarly, Eqs. (\ref{eq9}) can be Fourier-transformed over the transversal coordinates $(x,y)$, which leads to the following boundary conditions for the Fourier components of the transformed potential:
	\begin{subequations}
		\begin{align}
		&\dfrac{\partial \hat{\psi}(\bm{q},z)}{\partial z}\biggr\arrowvert_{z=(d/2)^+}-\dfrac{\partial \hat{\psi}(\bm{q},z)}{\partial z}\biggr\arrowvert_{z=(d/2)^-}=-4\pi\lambda_B\hat{\sigma}_1(\bm{q)},&\\
		&\dfrac{\partial \hat{\psi}(\bm{q},z)}{\partial z}\biggr\arrowvert_{z=(-d/2)^+}-\dfrac{\partial \hat{\psi}(\bm{q},z)}{\partial z}\biggr\arrowvert_{z=(-d/2)^-}  =-4\pi\lambda_B\hat{\sigma}_2(\bm{q}).&
		\end{align}
		\label{eq13}
	\end{subequations}
	The electrolytes in the outer-surface regions are in contact with an ionic reservoir of concentration $c_s$. If the ions in solution are monovalent, $\alpha_i=\pm 1$, positional correlations between them can be neglected, leading to a mean-field approximation $\rho_{\pm}(\bm{S},z)=c_se^{\mp\psi(\bm{S},z)}$ for the ionic profiles. In such a case, Eq. (\ref{eq8}) becomes the traditional mean-field Poisson-Boltzmann (PB) equation for the ionic distributions. In this situation, further progress requires the numerical integration of the PB equation, which is a non-trivial question when the charge distributions assigned to the charged surfaces are non-uniform~\cite{Sam19}. Instead of following this approach, we shall here consider a simplifying assumption which allows for exact solutions for the potential components -- namely the one of linearized ionic distributions. This assumption will be quite reasonable whenever the surface charges are not too high, in which case non-linear effects in the Stern layer close to surface contact can be safely neglected. As we will briefly see, this assumption leads to an exact integration of the averaged potential for both uniform and patterned-like surface distributions, which provides us valuable insights into the main physical properties that control EDLs interactions and charge neutrality in these systems. The validity of such physical mechanisms can be extended all the way to cases of non-linear surface-ion couplings, provided proper care is taken when incorporating the effects on non-linearity into the theory~\cite{levin}.

	According to the linear approximation, the density profiles of monovalent ions around the charged surfaces take the simple form $\hat{\rho}_{\pm}(\bm{q},z)=c_s[\delta(\bm{q})\pm\hat{\psi}(\bm{q},z)]$. Substitution of these profiles into Eq. (\ref{eq12}) results in the following Helmholtz equation for the Fourier components of the averaged electrostatic potential:
	\begin{equation}
	\dfrac{\partial^2\hat{\psi}(\bm{q},z)}{\partial z^2}=k^2\hat{\psi}(\bm{q},z).
	\label{eq14}
	\end{equation} 
	Here, the parameter $k$ is defined as $k=\sqrt{\kappa^2+q^2}$, where $\kappa\equiv \sqrt{8\pi\lambda_B c_s}$ is the traditional inverse Debye screening length, which sets up the screening of the surface charges by the surrounding electrolyte. Notice that the {\it overall} screening $k$ will now also depend on the inverse wavelength $q$, which in turn depends on the typical size of the charged domains on the surfaces. The boundary conditions to be enforced upon the potential components $\hat{\psi}(\bm{q},z)$ depend on whether the system boundaries are close of open.  We shall now consider separately the two distinct model systems for the confined electrolyte outlined in Section~\ref{model}. 
	
	\subsection{EXPLICIT RESERVOIR -- OSMOTIC EQUILIBRIUM}\label{m1}

	When the ionic particle reservoir is {\it explicitly} taken into account, ion exchange can take place between the confined electrolyte and the regions beyond the charged surfaces, in such a way that the electronically condition in the inter-surface region should not be taken as a {\it a priori} assumption. Instead, the system has to satisfy a  {\it global} electroneutrality condition. This can be achieved by considering the boundary conditions of vanishing electric fields far away from the charged surfaces,
	\begin{equation}
	\dfrac{\partial \hat{\psi}(\bm{q},z)}{\partial z}\biggr\arrowvert_{z\rightarrow -\infty}=\dfrac{\partial \hat{\psi}(\bm{q},z)}{\partial z}\biggr\arrowvert_{z\rightarrow \infty}=0.
	\label{eq15}
	\end{equation} 
	Apart from this condition, the electrostatic potential resulting from Eq. (\ref{eq14}) must also satisfy conditions (\ref{eq13}) due to the electric field provided by the charged surfaces. Note that the above conditions specify the asymptotic potential up to an arbitrary additive constant, which we here set to be zero, in such a way as to guarantee that the linearized ionic profiles $\rho_{\pm}(\bm{S},z)=c_s[1\mp \psi(\bm{S},z)]$ relax to their bulk (reservoir) values $c_s$ far away from the charged interface. Considering the general case where these surfaces are separated from the surrounding electrolyte by parallel neutral membranes located at  distances $z_m$ from the surfaces (see Fig.~\ref{fig:fig1}), the averaged potential has to additionally satisfy the Laplace equation $\nabla^2\psi(\bm{S},z)=0$ inside the regions $-l\le z \le -m$ and $m\le z \le l$, where $m\equiv d/2-z_m-r$ and $l\equiv d/2+z_m+r$ are the absolute values of the closest distance between the surface and the inside/outside electrolyte, respectively ($r=0.2$~nm is the ionic radii). Working in terms of the transversal Fourier components $\hat{\psi}(\bm{q},z)$, these conditions translate into the following differential equation for the averaged potentials:
	\begin{equation}
	\dfrac{\partial^2\hat{\psi}}{\partial z^2}(\bm{q},z)=\begin{cases}
	q^2\hat{\psi}(\bm{q},z),\hspace{1cm}|z\pm d/2|<s,\\
	k^2\hat{\psi}(\bm{q},z),\hspace{1cm}|z\pm d/2|>s,
	\end{cases}
	\label{eq01}
	\end{equation}
	where $s\equiv z_m+r$ defines the distance of closest ion-surface approach. Apart from boundary conditions (\ref{eq13}) and (\ref{eq15}), the potential components $\hat{\psi}(\bm{q},z)$ and their derivatives are further constrained to be continuous across the closest surface-ion distances $z=\pm l$ and $\pm m$. 
	
	Once the solution  $\hat{\psi}(\bm{q},z)$ of Eq. (\ref{eq01}) is found for the components  of the electrostatic potential, the osmotic pressure across the charged interfaces, defined as:
	\begin{equation}
	\beta\Pi\equiv-\dfrac{\partial\beta\mathcal{F}}{\partial V}=-\dfrac{1}{A}\dfrac{\partial\beta\mathcal{F}}{\partial d},
	\label{eq16}
	\end{equation} 
	where $A$ is the transversal surface area and $\mathcal{F}$ the system free-energy, can be readily evaluated. In fact, the osmotic pressure (\ref{eq16}) takes a particularly simple form when written in terms of the electric fields and density profiles induced at closest surface-ion contacts. The osmotic pressure between the interfaces comprises two contributions: one mechanical contribution $\Pi^{mec}$ due to the thermal collisions of the surrounding ions at the inner/outer interfaces and the electrostatic contribution $\Pi^{elec}$ resulting from both surface-ion and surface-surface electrostatic interactions. Explicitly, these contributions take the simple form:
	\begin{align}
	\beta\Pi & =  \beta\Pi^{mec}+\beta\Pi^{elec},\label{eq17}\\
	\beta\Pi^{mec} & = \dfrac{(2\pi)^2}{2A}\left[\delta\hat{\rho}_1(\bm{q}=0)-\delta\hat{\rho}_2(\bm{q}=0)\right],\label{eq19}\\
	\beta\Pi^{el} & =\dfrac{(2\pi)^2}{2A}\Biggl[\int\hat{\sigma}_2(\bm{q})\left(\hat{E}_z(-\bm{q},d/2)-\hat{E}_z^{(2)}(-\bm{q},d/2)\right)d\bm{q}\nonumber\\
	&-\int\hat{\sigma}_1(\bm{q})\left(\hat{E}_z(-\bm{q},-d/2)-\hat{E}_z^{(1)}(-\bm{q},-d/2)\right)d\bm{q}\Biggl].\label{eq20}
	\end{align}
	Here, $\hat{E}_z\equiv -\frac{\partial \hat{\psi}}{\partial z}$ is the longitudinal component of the total electric field, while $\hat{E}^{(1)}_z\equiv -\frac{\partial \hat{\phi}_1}{\partial z}$ and $\hat{E}^{(2)}_z\equiv -\frac{\partial \hat{\phi}_2}{\partial z}$ are the $z$-component of the electric fields produced by the charged surfaces $1$ and $2$, respectively, with corresponding potentials $\hat{\phi}_1(q)$ and $\hat{\phi}_2(q)$ given by:
	\begin{subequations}
		\begin{eqnarray}
		\hat{\phi}_1(\bm{q},z) & = & \dfrac{2\pi\lambda_B\hat{\sigma}_1(\bm{q})}{q}
		\begin{cases}
		e^{q(z+d/2)},\hspace{1cm} z<-d/2,\\
		e^{-q(z+d/2)},\hspace{0.85cm} z\ge-d/2.\\
		\end{cases}
		\\
		\hat{\phi}_2(\bm{q},z) & = & \dfrac{2\pi\lambda_B\hat{\sigma}_2(\bm{q})}{q}
		\begin{cases}
		e^{q(z-d/2)},\hspace{1cm} z<d/2,\\
		e^{-q(z-d/2)},\hspace{0.85cm} z\ge d/2.
		\end{cases}
		\end{eqnarray}
	\end{subequations}
	
	In Eq. (\ref{eq19}), $\delta\hat{\rho}_1(\bm{q})\equiv\hat{\rho}_1(\bm{q},-m)-\hat{\rho}_1(\bm{q},-l)$ and $\delta\hat{\rho}_2(\bm{q})=\hat{\rho}_1(\bm{q},l)-\hat{\rho}_1(\bm{q},m)$ are the ionic density differences across their inner and outer closest contact approaches with walls $1$ and $2$, respectively. Clearly, this mechanical contribution is due to momentum transfer from the ionic species at the wall surfaces. On the other hand, the electric contribution $\beta\Pi^{el}$ in Eq. (\ref{eq20}) represents the total electrostatic force on the charged walls (excluding a spurious self-interaction contribution). The above relations are exact, and do not rely on the particular approximation model implemented to compute the ionic profiles and electric field. A detailed derivation of these equations using a thermodynamic route is provided in \ref{Appendix1}.
	
	Another important quantity that can be readily obtained from the solutions of Eq. (\ref{eq14}) is the degree of electroneutrality in the region confined between the charged surfaces. The total ionic charge density confined into this region is:
	\begin{align}
	\sigma_{ion}\equiv\dfrac{1}{A}\sum_{i=\pm}\int_{-m}^{m}dz\int\alpha_i\rho_i(\bm{S},z)d\bm{S}\nonumber\\=\dfrac{(2\pi)^2}{A}\sum_{i=\pm}\alpha_i\int_{-m}^{m}\hat{\rho}_i(\bm{q}=0,z)dz.
	\label{eq21}
	\end{align}
	For the case of monovalent ions of bulk concentration $c_s$ in the context of the Debye-H\"uckel (DH) approximation $\hat{\rho}_{\pm}(\bm{q},z)=c_s\left[\delta(\bm{q})\mp\hat{\psi}(\bm{q},z)\right]$, the expression above simplifies to:
	\begin{equation}
	\sigma_{ion}=-\dfrac{\pi\kappa^2}{\lambda_BA}\int_{-m}^{m}\hat{\psi}(\bm{q}=0,z)dz. 
	\label{eq22}
	\end{equation}
	Notice that, due to the linearity of the averaged potential, $\hat{\psi}(\bm{q},z)$ will be directly proportional to the surface charge densities $\hat{\psi}(\bm{q},z)\sim\hat{\sigma}_{1}(\bm{q})$ and $\hat{\psi}(\bm{q},z)\sim\hat{\sigma}_{2}(\bm{q})$. Therefore, the relation above makes clear that only the {\it monopole} contribution $\hat{\sigma}(\bm{q}\rightarrow 0)$ will effectively contribute to the surface charge assign to the confined electrolyte. In particular, if the monopole ({\it i. e.}, the net charge) on the surfaces vanishes, $\hat{\sigma}(\bm{q}\rightarrow 0)=0$, electroneutrality will not be violated in the inter-surface space. It is thus convenient to represent the inhomogeneous surface charge densities as a combination of a monopole, uniform charge distribution $\hat{\sigma}_{h}(\bm{q})=\sigma_0\delta(\bm{q})$, plus a multi-pole charge contribution $\hat{\sigma}_{m}(\bm{q})$ bearing zero net charge, $\hat{\sigma}_{m}(\bm{q}\rightarrow 0)=0$. Clearly, the monopole contribution arising from $\hat{\sigma}_{h}(\bm{q})$ will be the dominant contribution to the electrostatic interactions, the remaining  multipole contributions being stored altogether in the inhomogeneous distribution $\hat{\sigma}_{m}(\bm{q}\rightarrow 0)$. In this work, $\hat{\sigma}_{h}(\bm{q})$ stands for the background surface charge density in Eq.(\ref{eq1}), whereas  $\hat{\sigma}_{m}(\bm{q})$ is represented by the modulated sinusoidal charge distributions, which in Fourier space factorizes as: 
	\begin{align}
	\hat{\sigma}(\bm{q})=\dfrac{\sigma_0}{(2i)^2}\left[e^{i\varphi_{x}}\delta(q_{x}-K_{x})-e^{-i\varphi_{x}}\delta(q_x+K_{x})\right]\left[e^{i\varphi_{y}}\delta(q_{y}-K_{y})-e^{-i\varphi_{y}}\delta(q_y+K_{y})\right]. 
	\label{eq23}
	\end{align}
	
	Once the Fourier components of the averaged potential $\hat{\psi}(\bm{q},z)$ are computed by solving Eq. (\ref{eq14}) with the proper boundary conditions, Eqs. (\ref{eq17}), (\ref{eq19}), (\ref{eq20}), and (\ref{eq22}) can be used to access the induced interaction and the electroneutrality condition in the inter-surface space. Explicit expressions for the solutions $\hat{\psi}(\bm{q},z)$ in the case of arbitrary surface charge modulations and ion-surface closest separations are shown in ~\ref{Appendix2}. Here we only show the expressions for the osmotic pressure and confined ionic charge. Since the linear approximation always predicts the same contact ionic densities in both sizes of the membrane, the mechanical contribution, Eq. (\ref{eq19}), always vanishes in the framework of DH theory. Another direct consequence of linearization is the absence of coupled monopole-multipole interactions, as can be readily verified from Eq. (\ref{eq20}) by noticing that the $\bm{q}\rightarrow 0$ mode from the multipole field vanishes. As a result, the electrostatic contribution to the osmotic pressure can be written as a simple combination of pure monopole and multipole contributions. The contribution from the uniform, monopole charge charge  can be written as: 
	\begin{equation}
	\beta\Pi^{el}_h=\zeta^0_1(\sigma_{01}^2+\sigma_{02}^2)+2\zeta^0_2\sigma_{01}\sigma_{02},
	\label{eq24}
	\end{equation}   
	where $\sigma_{01}$ and $\sigma_{02}$ represent the monopole charges on plates $1$ and $2$, respectively. The coefficients $\zeta^0_1$ and $\zeta^0_2$ defined above are given by:
	\begin{eqnarray}
	\zeta^0_1 =\pi\lambda_B\left(\dfrac{1+\kappa se^{-2\kappa m}}{(1+\kappa s)^2e^{2\kappa m}+(\kappa s)^2e^{-2\kappa m}}\right),\label{eq25a}\\
	\zeta^0_2=\pi\lambda_B\left(\dfrac{1+\kappa s}{(1+\kappa s)^2e^{2\kappa m}+(\kappa s)^2e^{-2\kappa m}}\right)\label{eq25b},
	\end{eqnarray}
	where $s\equiv r+z_m$ is the closest wall-ion distance. Since the coefficients above are always positive, the first contribution in Eq. (\ref{eq24}) is always repulsive, regardless the sign of the wall surface charges, while the second contribution is repulsive (attractive) depending whether the surfaces are equally (oppositely) charged. At large surface-surface separations $m\gg 1/\kappa$, the coefficient $\zeta^0_1$ decays to zero as $\sim e^{-4\kappa m}$, whereas $\zeta^0_2$ displays a slowly decay like $\sim e^{-2\kappa m}$, so that the second term in Eq. (\ref{eq24}) is the leading asymptotic contribution to the EDLs interactions. As expected, the effective surface interactions scale as $\sim\hat{\sigma}_1\hat{\sigma}_2 e^{-\kappa d}$ at large separations, reflecting the screening of the monopole interactions by the confined counterions.
	
	The multipole  contribution $\Pi^{el}_m$ to the osmotic pressure has a structure similar to Eq. (\ref{eq24}), but is now a superposition of all non-vanishing $\bm{q}$ modulations, that is,
	\begin{align}
	\beta\Pi^{el}_m  =  \dfrac{(2\pi)^2}{A}\Biggl[\int\zeta_1(q)\left(\hat{\sigma}_{1m}(\bm{q})\hat{\sigma}_{1m}(-\bm{q})+\hat{\sigma}_{2m}(\bm{q})\hat{\sigma}_{2m}(-\bm{q})\right)d\bm{q}\nonumber\\
	+\int\zeta_2(q)\left(\hat{\sigma}_{1m}(-\bm{q})\hat{\sigma}_{2m}(\bm{q})+\hat{\sigma}_{1m}(\bm{q})\hat{\sigma}_{2m}(-\bm{q})\right)d\bm{q}\Biggl],
	\label{eq26}
	\end{align} 
	where now $\hat{\sigma}_{1m}(\bm{q})$ and $\hat{\sigma}_{2m}(\bm{q})$ denote the $\bm{q}$-components of the multipole surface charge distributions. The coefficients $\zeta_1(q)$ and $\zeta_2(q)$ are explicit functions of the charge modulation $q$, and have the general form:
	\begin{align}
	\zeta_1(q) &=  \dfrac{2\pi\lambda_B}{\Delta}kq(k^2-q^2)\sinh(2qs)e^{-2km},\label{eq27}\\
	\zeta_2(q) & =  \dfrac{\pi\lambda_Bkqf(2qs)}{4\Delta}.\label{eq28}
	\end{align}
	In the above relations, we have introduced a function $f(x)$ defined as:
	\begin{equation}
	f(x)\equiv (k^2+q^2)\sinh(x)+2qk\cosh(x),
	\label{eq29}
	\end{equation}
	as well as the parameter $\Delta$, which reads as:
	\begin{equation}
	\Delta\equiv(k+q)^2\sinh^2(2qs)f(2km)+2qk\left(f(2qs)-qke^{-2qs}\right)e^{2(km-qs)}.
	\label{Delta}
	\end{equation}
	
	Notice that the coefficients $\zeta^0_1$ and $\zeta^0_2$ in Eqs. (\ref{eq25a}) and (\ref{eq25b}) can be obtained from Eqs. (\ref{eq27}) and (\ref{eq28}) by taking the limit $q\rightarrow 0$. The specific behavior of the osmotic pressure $\beta\Pi^{el}_m$ in (\ref{eq26}) as a function of the surface separation $d$ depends on the inhomogeneous charge distributions $\hat{\sigma}_1(\bm{q})$ and $\hat{\sigma}_2(\bm{q})$ over the surfaces. For the surface charges with periodic stripe-like modulations introduced in Eq. (\ref{eq1}), replacement of Eq. (\ref{eq23}) for the surface charge densities into (\ref{eq26}) results in the following multipole contribution to the electrostatic pressure:
	\begin{equation}
	\beta\Pi^{el}_m = \zeta_1(K)\left({\sigma}_{01}^2+{\sigma}_{02}^2\right)+2\zeta_{2}(K){\sigma}_{01}{\sigma}_{02}\cos(\delta\varphi_x)\cos(\delta\varphi_y)\delta_{\bm{K}_1\bm{K}_2}.
	\label{eq30}
	\end{equation}
	Here, $\bm{K}\equiv K_x\hat{\bm{e}}_x+K_y\hat{\bm{e}}_y$ is the wavenumber vector assign to the charge modulations,  $\delta\varphi_x\equiv\varphi_{x1}-\varphi_{x2}$ and $\delta\varphi_y\equiv\varphi_{y1}-\varphi_{y2}$ are phase differences between the charge stripes in the two surfaces across the $x$ and $y$ directions, respectively. The coefficients $\zeta_1(q=K)$ and $\zeta_1(q=K)$ are both positive, and display a similar behavior with respect to the surface separation $d$ as the coefficients $\zeta^0_1$ and $\zeta^0_2$ in Eq. (\ref{eq24}). The difference is that now the screening constant which dictates the exponential decay at large surface separations, ($\zeta_1\sim e^{-2kd}$ and  $\zeta_2\sim e^{-kd}$) is $k=\sqrt{\kappa^2+K^2}>\kappa$.  As expected, the multipole contributions will decay faster than their monopole counterparts. Moreover, the long range decay is inversely proportional to the periodic size domains $L_x$ and $L_y$. As the charged stripes become thinner, the resolution of the fine details of the inhomogeneous charge distributions becomes increasingly weaker at large distances, such that the inhomogeneous charge field rapidly resembles that of a locally neutral surface charge. Another difference with respect to the monopole case is that the sign of the coupled interaction in Eq. (\ref{eq30}) is now dictated by the phase differences on the domain distributions across the $x$ and $y$ directions. As the phase difference $\delta\varphi$ in any direction changes, the coupled surface-surface interactions in Eq. (\ref{eq30}) continuously interpolate from repulse to attractive. In particular, when the charged domains in both walls are completely out of phase ($\delta\varphi=\pi/2$), the net coupled force will be zero. This is the reason why decorating surfaces with charged patchy-like domains has attracted growing attention as promising strategy for tuning the induced interactions between charged surfaces~\cite{Bia14,Bia17}. 
	
	The electroneutrality degree into the confined region can be analyzed by inserting the obtained $q\rightarrow 0$ mode of the averaged potential into Eq. (\ref{eq22}). The result is:
	\begin{equation}
	\sigma_{ion}=-\dfrac{(\sigma_{01}+\sigma_{02})}{2}\dfrac{(\kappa s+1)\left[(\kappa s+1)e^{2\kappa m}-\kappa se^{-2\kappa m}-1\right]}{(\kappa s+1)^2e^{2\kappa m}-(\kappa s)^2e^{-2\kappa m}}.
	\label{eq31}
	\end{equation}
	One way of quantify the lack of electroneutrality in the confined electrolyte ($-m\le z\le m$) is by defining a net surface charge within this region as:~\cite{colla} 
	\begin{equation}
	\Gamma\equiv\dfrac{(\sigma_{01}+\sigma_{02})}{2} + \sigma_{ion}.
	\label{eq32}
	\end{equation}
	Notice that only half of monopole surface charge is included in this definition. This is because only half of the surface charge will be facing the confined electrolyte ({\it i. e.}, at the inner surface), while its other side will be in the outer face, facing towards the ionic reservoir. When local charge neutrality is satisfied, the ionic mean charge density $\sigma_{ion}$ balances the inner monopole surface charge, and $\Gamma$ approaches zero. Since the confined region can not be overcharged, $\Gamma$ has the same sign as the net monopole surface charge. Substitution of Eq. (\ref{eq31}) into Eq. (\ref{eq32}) leads to
	\begin{equation}
	\Gamma = \dfrac{(\sigma_{01}+\sigma_{02})}{2}\left(\dfrac{\kappa se^{-2\kappa m}+1}{(\kappa s+1)^2e^{2\kappa m}-(\kappa s)^2e^{-2\kappa m}} \right).
	\label{eq32}
	\end{equation}
	At large surface separations, the leading decay of $\Gamma$ goes as $\sim e^{-\kappa d}$. At high bulk salt concentrations $c_s$, ions will diffuse easily into the confining region, and electroneutrality will set up at smaller surface separations $d$. When the salt concentration is small, the confined ions will not be able to fully screen the inner surface charge and local charge neutrality will take place at smaller wall separations. Before considering the effects of such local electroneutrality breakdown on the induced surface interactions, it is instructive to consider a second model system in which charge neutrality is {\it enforced} in the inter-surface region.
	
	\subsection{IMPLICIT RESERVOIR -- DONNAN EQUILIBRIUM}\label{m2}

	We now consider the second model system, depicted in Fig.~\ref{fig:fig1}b, in which the confined electrolyte is connected to an implicit ionic reservoir of concentration $c_s$. The region beyond the surfaces is thus covered by an inert electrolyte of zero local net charge and same dielectric constant. This condition implicit requires that the monopole electric field has to be confined in the inter-surface region. In other words, charge neutrality is enforced as a boundary condition for the confined electrolyte. In practice, such requirement is achieved {\it via} the building-up of a potential difference across the system-reservoir interface, which prevents the monopole contribution to the electric field to leak out of the system boundaries \footnote {Notice that we do not enforce the multipole contributions $q\ne 0$ to the electric field to be also confined in the inter-surface space. Despite this being not a necessary condition on physical grounds, alternative Donnan approaches might impose such extra constrain for the multipole electric field as well.}. This potential is widely known as the {\it Donnan potential}, and it is in practice implicitly incorporated into the theoretical description through the boundary condition of vanishing monopole fields at the system interface. 
	
	In order to study the effects of Donnan equilibrium in the surface interactions, we again employ the linear approximation for the  averaged potential, which must now satisfy the Laplace equation $\nabla^2\psi=0$ for $|z|>m$ and the Helmholtz equation, $\nabla^2\psi=k^2\psi$ in the space between the surfaces, $|z|<m$. After a Fourier transformation over the transversal coordinates, these equations become:
	\begin{equation}
	\dfrac{\partial^2\hat{\psi}}{\partial z^2}(\bm{q},z)=\begin{cases}
	q^2\hat{\psi}(\bm{q},z),\hspace{1cm}|z|>m\\
	k^2\hat{\psi}(\bm{q},z),\hspace{1cm}|z|\le m.
	\end{cases}
	\label{eq33}
	\end{equation}
	These equations have to be solved subject to the boundary conditions of vanishing field at infinity, Eq. (\ref{eq15}). Since the monopole charges vanishes identically in the region beyond the surfaces, this condition implies that the monopole contribution to the electric field has to be zero just outside the confined electrolyte. The zeroth moment averaged potential $\hat{\psi}(\bm{q}\rightarrow 0,z)$ is thus bound to satisfy the following boundary condition at the interface:
	\begin{subequations}
		\begin{align}
		\dfrac{\partial\hat{\psi}(\bm{q}\rightarrow 0,z)}{\partial z}\Biggr\arrowvert_{z=-d/2} & =-4\pi\sigma_{01},\label{eq34a}\\
		\dfrac{\partial\hat{\psi}(\bm{q}\rightarrow 0,z)}{\partial z}\Biggr\arrowvert_{z=d/2} & =4\pi\sigma_{02}.\label{eq34b}
		\end{align}
	\end{subequations}
	The chemical equilibrium with a neutral reservoir of concentration $c_s$ is now implicitly assumed through the ionic profiles $\rho_{\pm}(\bm{S},z)=c_s\exp(\pm\psi(\bm{S},z))\approx c_s[1\mp\psi(\bm{S},z)]$, which now has to be supplemented with a local electroneutrality condition. Explicit expressions for the solution of Eqs. (\ref{eq33}) subject to the zeroth-moment conditions (\ref{eq34a}) and (\ref{eq34b}) are shown in \ref{Appendix2}. 
	
	We now proceed to compute the osmotic stress between the surfaces in this model system. It is however important to note that  Eqs. (\ref{eq19}) and (\ref{eq20}) applied previously for computing the osmotic stress in the case of an explicit reservoir {\it do not hold} in the present situation. Why is that? This is because the Hamiltonian considered in the previous case has only an explicit dependence on the surface separation $d$ through the surface-ion and surface-surface interactions (see \ref{Appendix1}). This is no longer the case here, since there will be now an extra $d$ dependence ``hidden" in the Hamiltonian due to the charge neutrality constrain,
	\begin{equation}
	\sum_{i=\pm}\alpha_i\int_{-m}^{m}\hat{\rho}_i(\bm{q}\rightarrow 0,z)dz+\hat{\sigma}_1(\bm{q}\rightarrow 0)+\hat{\sigma}_2(\bm{q}\rightarrow 0)=0,
	\label{neutrality}
	\end{equation}
	which is implicitly assumed in Eqs. (\ref{eq34a}) and (\ref{eq34b}). This additional contributions comes from the Donnan potential which will set in at the interface. Equations (\ref{eq19}) and (\ref{eq20}) thus only provide the {\it ionic induced surface force}, but not the {\it osmotic stress} on these surfaces. From a thermodynamic perspective, the Free Energy in Eq. (\ref{eq16}) will also comprise a $d$-dependent Lagrange Multiplier $\mu_D$ which will ensure overall that the charge neutrality condition (\ref{neutrality}) is fulfilled in the confined electrolyte~\cite{Tam98}. 
	
	In order to calculate the osmotic stress we shall therefore follow a different route. First, we apply a Kirkwood charging process to compute the change in Free Energy when the surfaces are adiabatically charged from zero up to their final charges. The result is~\cite{Mik94,Denton2007,Ben13}
	\begin{align}
	\dfrac{\beta\delta\mathcal{F}}{A} & =\dfrac{(2\pi)^2}{A}\int_0^1d\lambda\int\biggr[\hat{\sigma}_1(-\bm{q})\dfrac{\partial\hat{\psi}_{\lambda}}{\partial\lambda}(\bm{q},z=-d/2)+\hat{\sigma}_2(-\bm{q})\dfrac{\partial\hat{\psi}_{\lambda}}{\partial\lambda}(\bm{q},z=d/2)\biggr]d\bm{q},
	\label{eq35}
	\end{align}
	where $\lambda$ is a coupling parameter which scales linearly with the surface charges ($\hat\sigma(\bm{q})\rightarrow \lambda\hat\sigma(\bm{q})$), and $\hat{\psi}_{\lambda}(\bm{q},z)$ is the corresponding averaged potential. In the context of the employed linear approximation, the potential is linearly proportional to the surface charges (see \ref{Appendix2}), {\it i. e.} $\hat{\psi}_{\lambda}(\bm{q},z)=\lambda\hat{\psi}(\bm{q},z)$. The coupling integral above can thus be trivially calculated. Inserting above the solution of Eq. (\ref{eq33}) and performing the integration, we find a Free Energy change comprising a decoupled monopole and multipole interactions, $\delta\mathcal{F}=\delta\mathcal{F}_h+\delta\mathcal{F}_m$. The monopole interaction  has the form
	\begin{equation}
	\dfrac{\beta\delta\mathcal{F}_h}{A}=\chi^0_1(\sigma_{01}^2+\sigma_{02}^2)+2\sigma_{01}\sigma_{02}\chi_2^0,
	\label{eq36}
	\end{equation}
	with coefficients $\chi^0_1$ and $\chi^0_2$ given by
	\begin{align}
	\chi^0_1 & =  \dfrac{4\pi\lambda_B}{\kappa}\left[\dfrac{\cosh(2\kappa m)}{\sinh(2\kappa m)}+\kappa a\right],\label{eq37}\\
	\chi^0_2 & =  \dfrac{4\pi\lambda_B}{\kappa\sinh(2\kappa m)}\label{eq38}.
	\end{align}
	Considering the $K$-charge modulation of Eq. (\ref{eq23}), the multipole contribution $\delta\mathcal{F}_m$ becomes
	\begin{align}
	\dfrac{\beta\delta\mathcal{F}_m}{A}=\chi_1(K)(\sigma_{01}^2+\sigma_{02}^2)+2\sigma_{01}\sigma_{02}\chi_2(K)\cos(\delta\varphi_x)\cos(\delta\varphi_y)\delta_{\bm{K}_1}\delta_{\bm{K}_2},
	\label{eq39}
	\end{align}
	where the coefficients $\chi_1(q)$ and $\chi_2(q)$ can be written as
	\begin{align}
	\chi_1(q) & =  \dfrac{4\pi\lambda_B}{qf(2km)}e^{-qs}\left[f(2km)\cosh(qs)-g(2km)e^{-qs}\right]-\dfrac{2\pi\lambda_B}{q}\label{eq40}\\
	\chi_2(q) & =  \dfrac{4\pi\lambda_Bk}{f(2km)}e^{-2qs}\label{eq41}-
	\end{align}
	Here, $f(x)$ is the function defined in Eq. (\ref{eq29}), while $g(x)$ is defined as
	\begin{equation}
	g(x)\equiv k^2\sinh(x)+qk\cosh(x).
	\label{g}
	\end{equation}
	In the context of a linear response approximation, the first terms in Eqs. (\ref{eq36}) and (\ref{eq39}) can be assign to an ionic induced self-energy, while the second terms corresponds to the effective surface-surface interactions~\cite{Denton2007,Den99}. It is now a simple task to calculate the osmotic pressure using Eq. (\ref{eq16}), with the free-energy change obtained from Eqs. (\ref{eq36}) and (\ref{eq39}). The osmotic pressure also split into monopole and dipole contributions as in Eqs. (\ref{eq24}) and (\ref{eq30}), respectively. The zeroth moment coefficients now read as
	\begin{align}  
	\zeta^0_1 & = \dfrac{4\pi\lambda_B}{\sinh^2(2\kappa m)},\label{eq42}\\
	\zeta^0_2 & =  \dfrac{4\pi\lambda_B\cosh(2\kappa m)}{\sinh^2(2\kappa m)},\label{eq43}
	\end{align}
	while the multipole coefficients ($\bm{q}\ne 0$) take the form
	\begin{align}  
	\zeta_1(q) & = \dfrac{2\pi k^2(k^2-q^2)e^{-2qs}}{f^2(2km)},\label{eq44}\\
	\zeta_2(q) & = \dfrac{2\pi k^2qe^{-2qs}}{f^2(2km)}\left[(k^2+q^2)\cosh(2km)+2kq\sinh(2km)\right].\label{eq45}
	\end{align}
	Notice that, contrary to the previous model in which the coefficients in (\ref{eq24}) and (\ref{eq30}) remain finite at all surface separations, the coefficients above will diverge at the smallest inter-surface distance ($m\rightarrow 0$). This singularity can be attributed to a divergence of the Donnan potential, as the reservoir would have to perform an infinite amount of work against the surface fields in order to keep electroneutrality in the confined electrolyte. 
	
	\section{Results}\label{results}
	
	We start by analyzing the local violation of electroneutrality and its interplay with the surface interactions for the case of monovalent ions. In the presence of multivalent ions, the theory outlined above loses its validity, as the neglected ionic correlations become a relevant contribution. In this case, we will apply the simulation technique described in Section~\ref{model} to get insights on how ionic correlations and non-linear effects modify the simple scenario predicted by the linear approach. 
	
\begin{figure}[h!]
		\begin{center}
			\includegraphics[width=9cm,height=7cm]{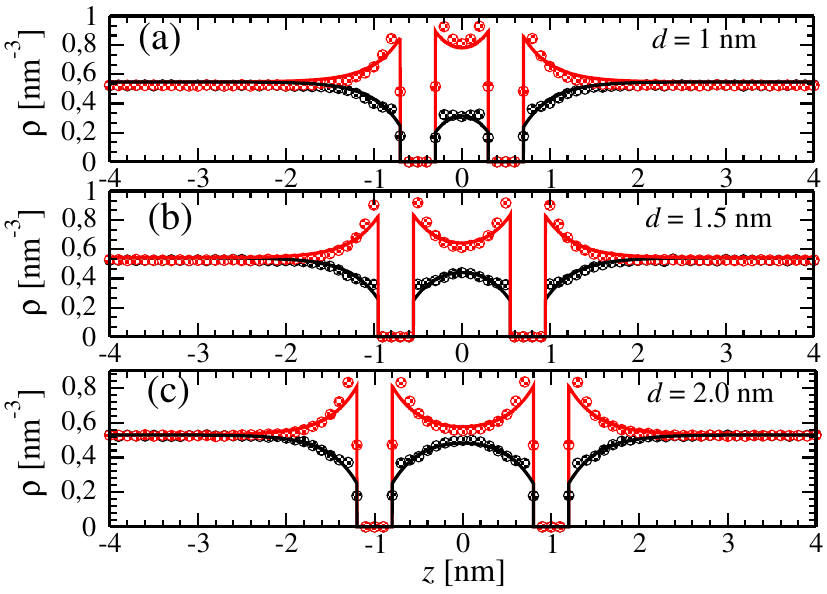}
		\end{center}
		\caption{Density profiles ions for $1:1$ salt for three difference separation distances. The concentration of salt is $75$ mM and  $\sigma_0 = 0.0575$ C/m$^2$. Symbols stand for simulation results, while solid lines are predictions for the linear approach.}  
		\label{fig:fig3}
	\end{figure}

	In Figure~\ref{fig:fig3}, averaged monopole ionic profiles $(2\pi)^2\hat{\rho}(\bm{q}=0,z)$ resulting from the DH model described in Section~\ref{m1} are compared with results from the MC technique outlined in Section~\ref{sim}. Notice that this quantity corresponds to the total charge per unit of area in a transversal plane located at a position $z$. Good agreement is observed between theory and simulations for all surface separations considered, in spite of a  tendency of the DH theory to slightly underestimate the contact densities at the surface.  The small discrepancies can be assigned to a coupling between monopole and multipole interactions, which turn out to be completely decoupled in the linear approach. Moreover, ionic size effects might become relevant at such strong confinements.

\begin{figure}
		\begin{center}
			\includegraphics[width=5.5cm]{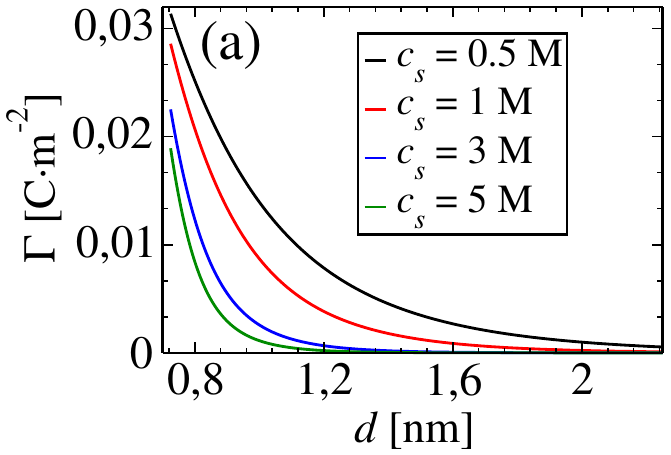}
			\includegraphics[width=5.5cm]{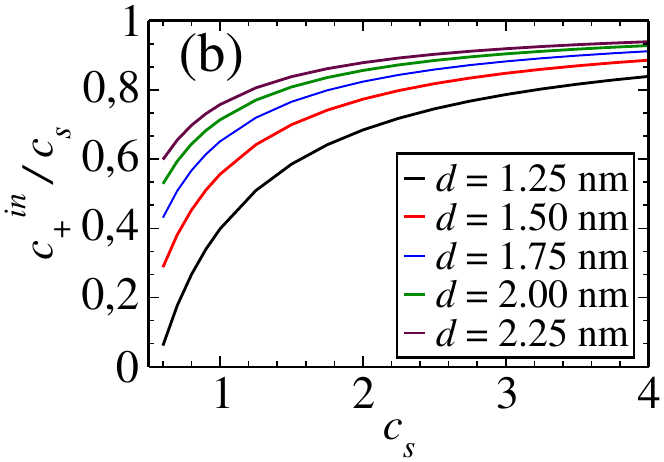}
		\end{center}
		\caption{Ionic penetration at the confined electrolyte. In (a), the confined net charge $\Gamma$ as a function of the surface distance for different ionic strengths is displayed. In panel (b), the ratio between co-ion concentrations inside ($c_s^{in}$) and outise ($c_s$) the confining region is shown for different inter-surface separations $d$. In both cases, the ionic radii is $r_i=0.2$~nm, the surface thickness is $z_m=0.25$~nm, and the surface charges are $\sigma_{01}=\sigma_{02}=0.075$~C/m$^2$.}
		\label{fig:fig4}
	\end{figure}

	Figure \ref{fig:fig4}a depicts the general behavior of the confined net charge $\Gamma$ calculated from Eq. (\ref{eq32}) at different ionic strengths. Local electroneutrality violation takes place at very short inter-surface separations. The ionic distributions are dictated by a balance of electrostatic contributions, which attempts to establish a local charge neutrality all over the system, and entropic effects, which favor homogeneous particle local densities everywhere. Due to the strong confinement, entropic effects prevents strong ionic condensation in the narrow inter-plate region, and the confined electrolyte is unable to keep its charge neutrality. As the surface separation increases, the ionic flux through the semi-permeable membranes rapidly restores local electroneutrality. This effect clearly depends on the bulk ionic concentrations. At high salt concentrations, the entropic cost for local ionic inhomogeneities is decreased. This means that ions will be able to pack more efficiently at the vicinity of charged surfaces in order to neutralize their charges, leading to a strong screening of electrostatic interactions. When the reservoir ionic concentration is decreased, entropic effects become more relevant, thus preventing a strong ionic packing at small regions~\cite{Col14}. The double layers become more diffuse in this limit, as ions need to rearrange into a larger distance in order to fully screen the surface charges. A measurement of the typical distance in which fully screening takes place is provided by the Debye screening length, $\kappa^{-1}$. Therefore, local electroneutrality will occur naturally when the inter-surface distance $d$ is comparable to the Debye length, $d\sim\kappa^{-1}$.     
	
	The interplay between electrostatic and entropic contributions in the confined electrolyte can also be measured by the so-called partitioning coefficient of co-ions, which is defined as the ratio between the mean concentration of this component in its confined ($c_{+}^{in}$) and bulk ($c_s$) phases \cite{Gou00,Gou01,Qiao18}. When electroneutrality is not achieved in the inter-surface space, co-ions will undergo an overall repulsion when entering this region. As a result of such electrostatic penalty, the overall co-ion concentration in the confined electrolyte will be much smaller than that of the bulk reservoir. On the other hand, as the ionic bulk concentration increases and/or the inter-surface distance becomes larger, entropic contributions will favor equal co-ion concentrations both inside and outside the inter-surface region. These trends are clearly observed in Fig.~\ref{fig:fig4}b, in which the partitioning coefficient of co-ions is displayed as a function of the reservoir concentration $c_s$ for several inter-plate distances. We notice that these results are in qualitative agreement with predictions of Ref. \cite{Jar09}, in which ionic correlations and solvent effects are incorporated in a coarse-graining approach combining both Molecular and Brownian Dynamics simulations. The smaller values of the fraction $c_+^{in}/c_s$ obtained in Ref. \cite{Jar09} indicates that the linear approach tend to underestimate the degree of electroneutrality, which should be further enhanced by inclusion of size and correlation effects.
	
\begin{figure}[h!]
		\begin{center}
			\includegraphics[width=7cm]{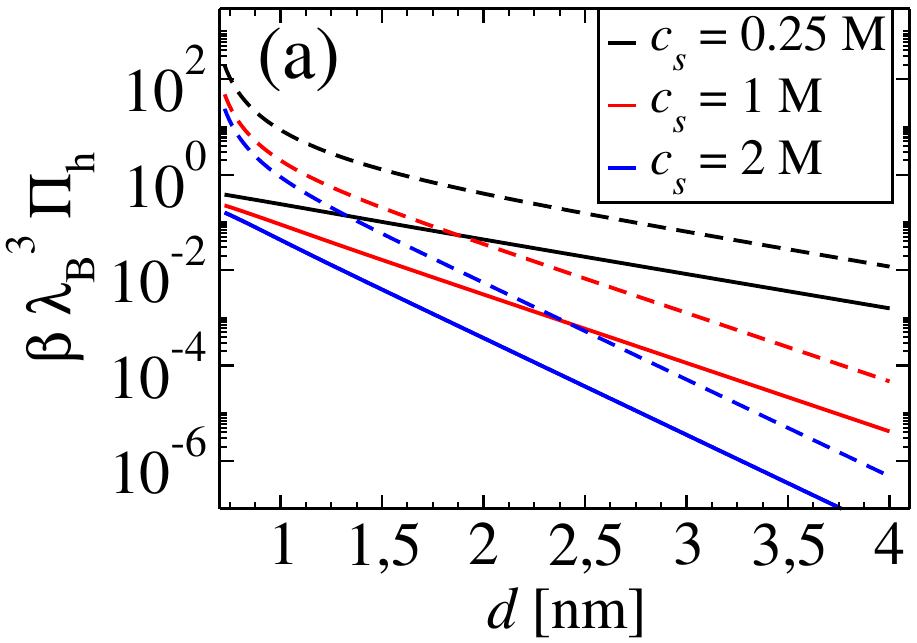}
			\includegraphics[width=7cm]{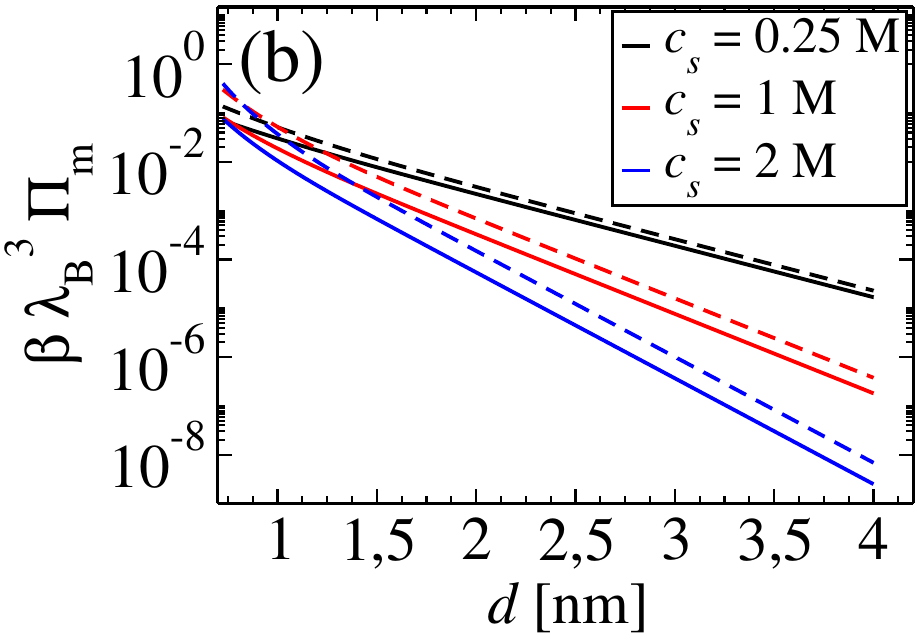}
		\end{center}
		\caption{Osmotic stress as a function of surface separation for equally charged surfaces bearing charge $\sigma_{01}=\sigma_{02}=0.075$~C/m$^2$. The surface thickness is  $z_m=0.25$~nm, and the ionic radii $r=0.2$~nm, (implying in a closest ion-surface distance of $s=0.45$~nm). Dashed lines stand for the predictions of implicit reservoir (Donnan model), while solid lines represent results from the explicit-reservoir model. In (a), the monopole contributions to the pressure are shown, whereas results from the multipole charge modulations are displayed in (b).}  
		\label{fig:fig5}
	\end{figure}

	The physical picture described above is in strong contrast with the one underlying a Donnan equilibrium with the charge reservoir. In that case, the confined electrolyte will always achieve charge neutrality, as the reservoir is able to provide an arbitrary large amount of neutralizing counterions. At strong confinements, an implicit Donnan potential is established across the interface in order to fulfill this requirement. Moreover, the absence of net monopole charges beyond the surfaces imply that EDLs are built up only in the inner faces of each surface. In contrast, the EDLs will be distributed over both inner and outer faces of the surfaces, when ions are freely allowed to diffuse between these regions. Clearly, these two distinct scenarios of ionic confinement -- free osmotic or Donnan-like equilibrium with the external environment -- will lead to different  induced interactions between  the confining surfaces. This difference can be investigated by comparing the predictions for the osmotic stresses obtained from Eqs. (\ref{eq24}) and (\ref{eq30}) for both models. Such comparison is provided in Fig.~{\ref{fig:fig5}}, in which both monopole and multipole surface forces per area at different ionic strengths are shown. In all cases, the surfaces have equal monopole charges, $\sigma_{01}=\sigma_{02}\equiv\sigma_0=0.075$~C/m$^2$, and the phase shifts in both directions are zero, $\delta\varphi_x=\delta\varphi_y=0$. All the interactions are then purely repulsive. The $y$-axis is set in a logarithm scale to help visualization. 
	
	%%%%%%%%%%%%%%%%%%%%%%%%%%%%%%%%%%%% 

	All the osmotic pressures in Fig. \ref{fig:fig5} display an exponential decay at large surface separations. As the ionic concentration increases, the slope of the curves become more negative, reflecting the larger screening constant. Despite the similar decay, the pressures calculated in both models show important quantitative differences. The forces calculated from the Donnan model (dashed lines) are always stronger than the ones obtained by the explicit reservoir (full lines) model, the difference being more pronounced in the monopole case. Such enhanced stress in the Donnan model seems to have its roots on a fine balance between competing mechanisms. On one hand, the fact that the surface field in the Donnan picture is {\it confined} at the inter-surface space implies in a stronger electrostatic bare interaction among these surfaces, as compared to the explicit-reservoir case (strictly speaking, twice as large). On the other hand, in the explicit-reservoir model there will be an extra layer of condensed counterions at the {\it outer} surfaces, which will push them away from each other, contributing to an increase of the repulse interactions. The stronger interactions in Donnan model shown in Figure \ref{fig:fig5} indicates that the former contributions are dominant over the latter ones.
	
	At small wall separations, the Donnan osmotic pressure in Fig. \ref{fig:fig5} deviate from the simple exponential decay. This behavior can be assign to a non-linear increase in the Donnan potential at strong confinements. In fact, a close inspection into Eqs. (\ref{eq42}), (\ref{eq43}), (\ref{eq44}) and (\ref{eq45}) shows that the osmotic stress will diverge at closest surface contact ($d\rightarrow 2s$). In contrast, the osmotic pressures in the explicit-model case remains always finite, even at contact approach. This is clearly a manifestation of local electronically violation, which allows for a finite local concentrations, even at vanishing surface separations. 
	
\begin{figure}[h!]
		\begin{center}
			\includegraphics[width=7.8cm,height=5cm]{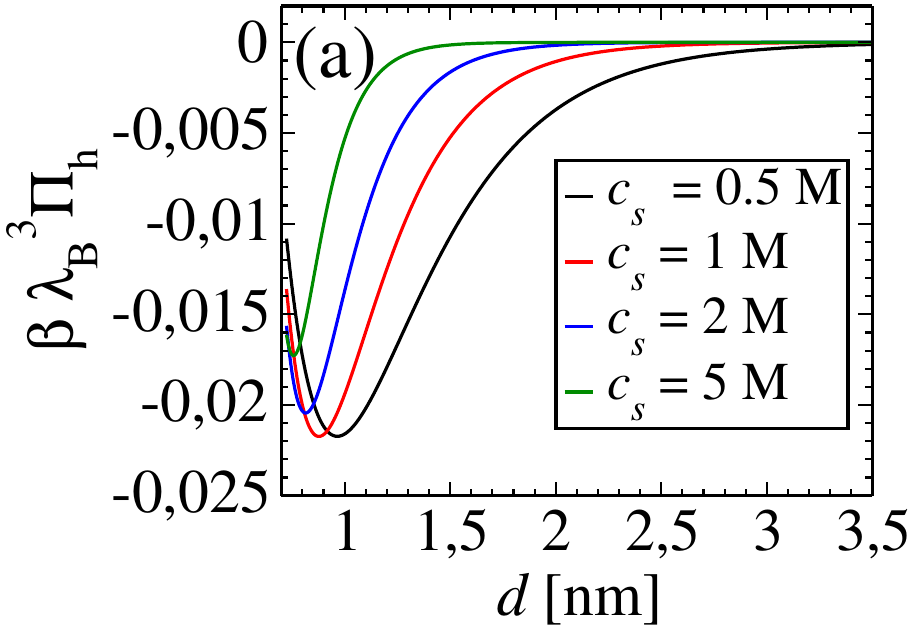}
			\includegraphics[width=7.5cm,height=5cm]{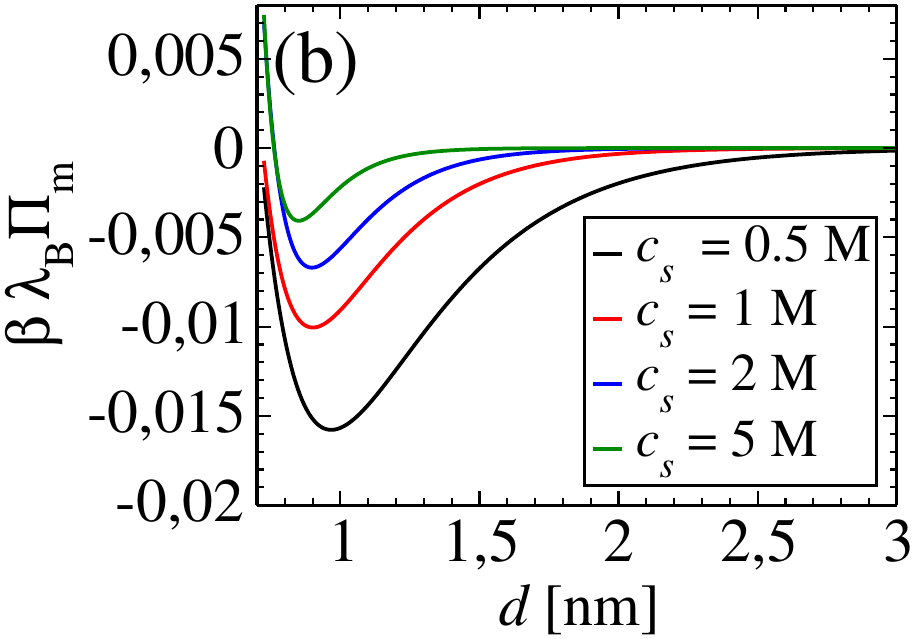}
		\end{center}
		\caption{Monopole (a) and dipole (b) contributions to the osmotic stress resulting from the explicit-reservoir model, in the case of oppositely charged walls of magnitude $\sigma_0=0.075$~C/m$^2$. The ionic radii is $r=0.2$~nm, while the surface thickness is $z_m=0.25$~nm, implying in a closest surface-ion distance of $s=0.45$~nm.}  
		\label{fig:fig6}
	\end{figure}

	Another interesting case is the one of oppositely charged surfaces with same magnitude, $\sigma_{01}=-|\sigma_{02}|$, and equal phase shifts $\delta\varphi_x=\delta\varphi_y$. The confined electrolyte will in this case always contain the same amount of negative and positive ions, and charge neutrality will be always satisfied, regardless the degree of confinement. This mechanism will avoid the emergence of a strong potential difference across the surface boundaries. Besides, it sets up a competition between repulsive and attractive interactions, which are in turn determined by the strengths of the coefficients $\zeta_1(q)$ and $\zeta_2(q)$ in Eq. (\ref{eq30}). Figs. \ref{fig:fig6}a and \ref{fig:fig6}b show the osmotic monopole and the multipole contributions to the pressures, respectively, obtained from the explicit-reservoir model. The osmotic pressures display in both cases a non-monotonic behavior, featuring a maximum attraction at small separations, followed by an exponential decay. This is clearly a consequence of a fine competition between an induced, repulsive self-energy (controlled by  $\zeta_1^0$ and $\zeta_1(q))$, and the attractive surface-surface interactions (standing from $\zeta_2^0$ and $\zeta_2(q)$). In the case of the case of the multipole interactions, these competing contributions lead to a crossover between pure attraction at small ionic concentrations and a short-range repulsive force at high ionic strengths and short separations (see Fig.~\ref{fig:fig6}b).
	
	%%%%%%%%%%%%%%%%%%%%%%%%%%%%%%%%%%%% 
	
\begin{figure}[h!]
		\begin{center}
			\includegraphics[width=7.8cm,height=5cm]{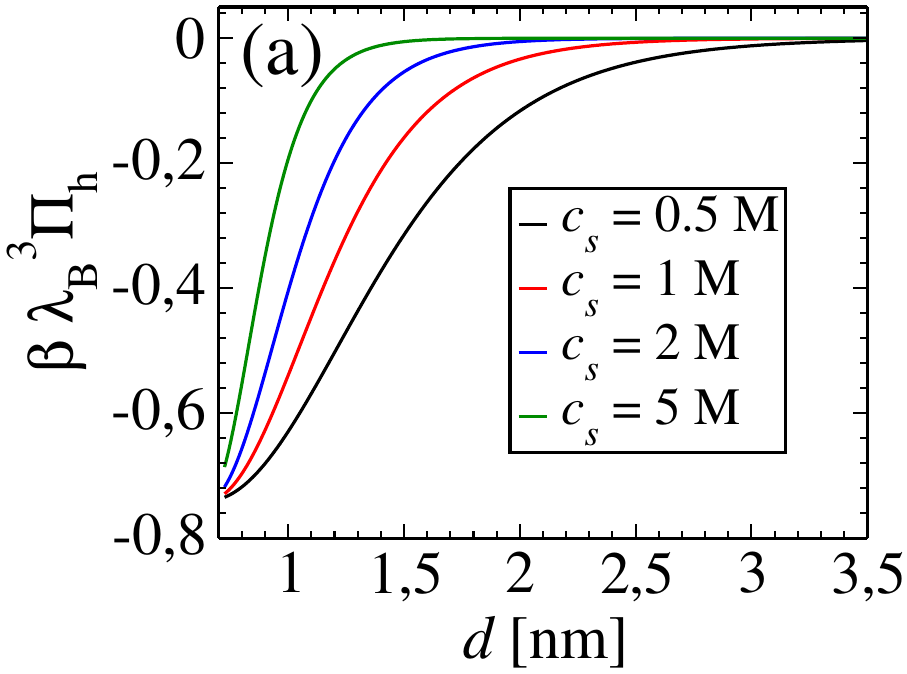}
			\includegraphics[width=7.8cm,height=5.2cm]{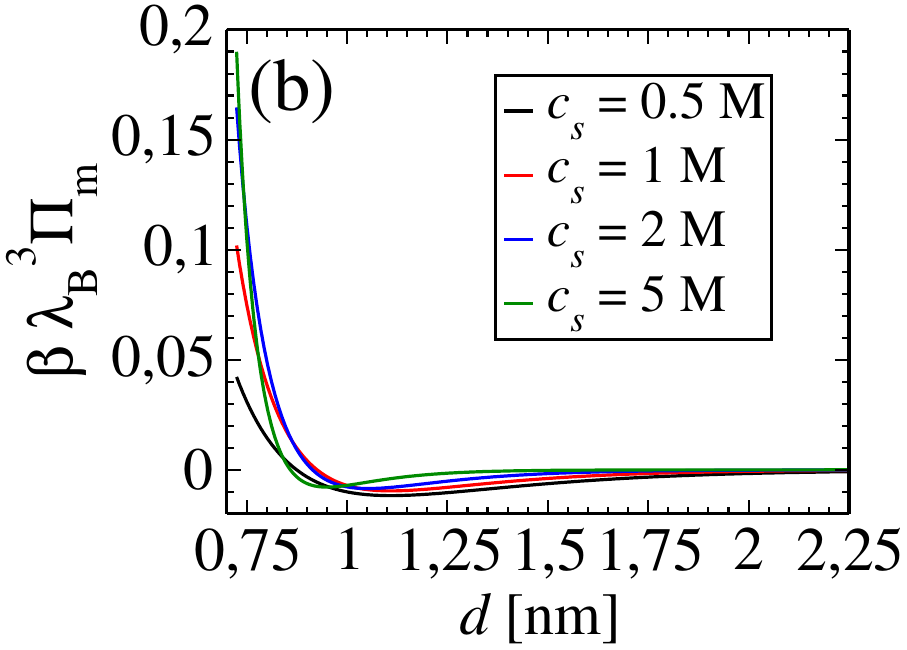}
		\end{center}
		\caption{Same as in Fig.~\ref{fig:fig6}, but with results obtained from the implicit-reservoir, Donnan model.}  
		\label{fig:fig7}
	\end{figure}

	A different qualitative behavior is predicted by the Donnan model, as shown in Figs. \ref{fig:fig7}a and \ref{fig:fig7}b for the monopole and multipole osmotic pressures, respectively. The monopole surface forces decay monotonically (in magnitude) at all observed salt concentrations. Furthermore, the strength of the induced surface attraction is remarkably larger in comparison to the implicit-reservoir model, Fig. \ref{fig:fig6}a. Again, this can be attributed to an additional contribution to the repulsive forces which comes from counterions condensed at the external sides of the surfaces, and which is absent in the Donnan approach. Another interesting point is the absence of diverging forces at close-contact separations, in strong contrast with the case of equally charged surfaces, see Fig. \ref{fig:fig5}a. This feature can be understood in terms of a lack of a Donnan potential, because the requirement of electronically is now naturally fulfilled.  We also notice that the multipole contributions in this case interpolate between a short range repulsion and a long-range attraction for all analyzed salt concentrations.

	%%%%%%%%%%%%%%%%%%%%%%%%%%%%%%%%%%%% 

	We now address the important question of the interplay between multipole contributions and the violation of local charge neutrality. Such interplay is absent in a linear treatment, since the the multipole contributions are always charge neutral, and fully decoupled from the monopole ones. This is no longer the case when effects such as non-linear contributions or ionic correlations are taken into account. It is clear that these effects will introduce a non-trivial coupling between monopole and multipole contributions. The addition of charge  modulations, for instance, in an otherwise uniformly charged surface will introduce a large number of extra neutralizing  ions into the system. At strong confinements, a large number of these ions will leave the confining region, which could effectively influence the monopole charge distributions and the underlying electroneutrality condition. Ionic correlations between ions condensed at the charged sites should also have non-trivial effects on charge neutrality and induced forces. In order to better understand the effects of the patterned charge surface on the electroneutrality violation, we have performed MC simulations at different charge modulations and inter-surface distances. We have also considered effects from ionic charge asymmetry. The results are summarized in Fig. ~\ref{fig:fig7}, in which  the confined net charge $\Gamma$ for three different surface distances are shown, considering distinct site modulations $(n_x,n_y)$. Overall, the results indicate that the inclusion of different modulations have a minor effect on the condensed charge in the case of monovalent ions. Small deviations from electroneutrality are observed at the shortest surface separation, indicating a weak degree of monopole-multipole couplings. The situation is changed in the presence of multivalent $2:1$ electrolyte, in which case ionic correlations become non-trivial. This leads to a stronger degree of charge neutrality violation, specially at short separations. In all cases, the electroneutrality is little influenced by the particular charge modulation. These results indicate that the overall physical picture outlined above for the case of weak couplings (namely weakly charged surfaces and negligible ionic correlations) could remain valid over a wider range of system parameters.

	\begin{figure}[h!]
		\begin{center}
			\includegraphics[width=5cm,height=3.7cm]{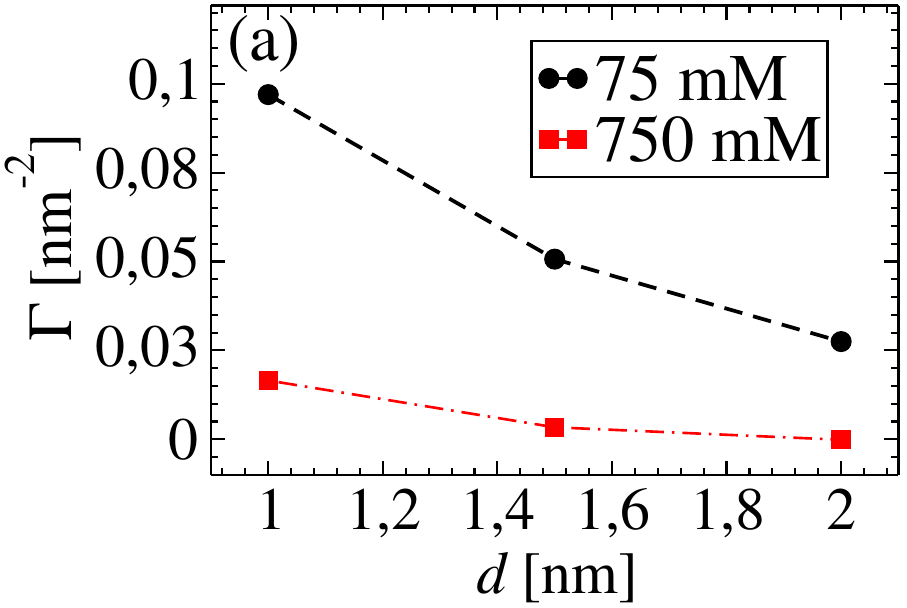}
			\includegraphics[width=5.2cm,height=3.7cm]{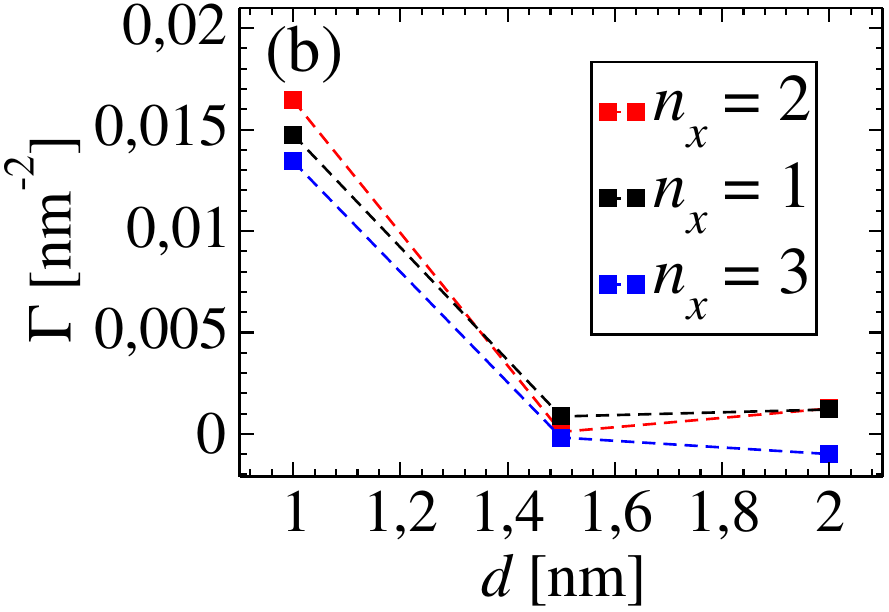}
			\includegraphics[width=5.2cm,height=3.7cm]{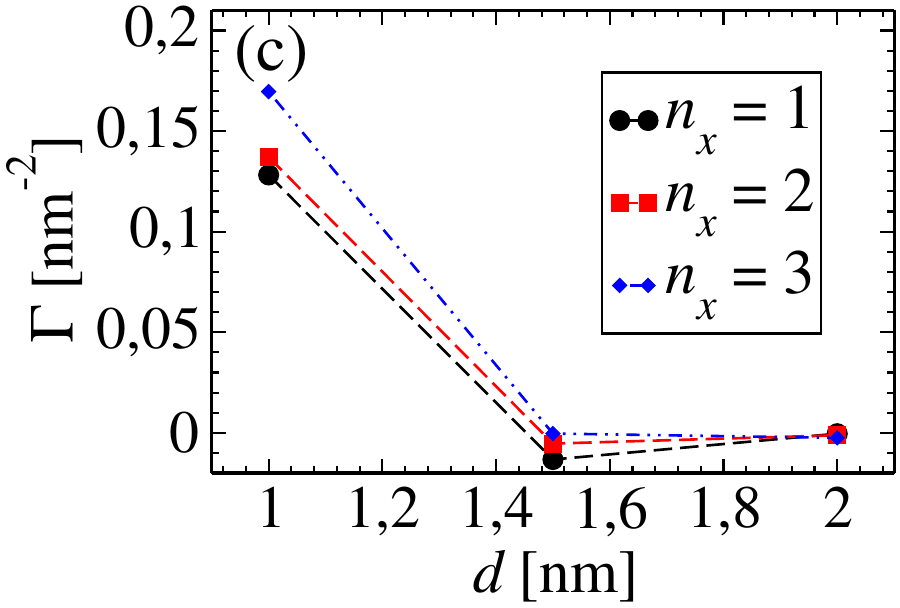}
		\end{center}
		\caption{$\Gamma$ for $1:1$ salt for three difference salt concentrations: $75$ mM (a), $750$ mM (b) and (c). The results in (a) and (b) are obtained using monovalent $1:1$ ions, while the predictions in (c) result from $2:1$ confined electrolytes. }  
		\label{fig7}
	\end{figure}

	\section{Conclusions}\label{conclusions}

	We have investigated the osmotic equilibrium and the surface interactions in the case of an electrolyte strongly confined between charged surfaces. The charges on the surfaces comprise a uniform background charge (monopole contribution) in addition to a patterned charge modulation with neutral net charge (multipole contributions). Two distinct models have been applied to investigate these systems: a model in which an osmotic equilibrium is established with an {\it explicit} ionic reservoir (allowing thus for the lack of local charge neutrality) and one in which electroneutrality is enforced {\it a priori} through the contact with an {\it implicit} reservoir (Donnan approach). It was shown that these models predict quite different behaviors for the induced monopole and multipole osmotic stresses, on both qualitatively and quantitatively grounds.  
	
	Our main conclusion is that the application of these two scenarios of osmotic equilibrium might lead to quite distinct behaviors. This point has to be taken into account when applying these models to calculate induced forces on charged interfaces. A common strategy to compute effective interactions between charged surfaces in solution is to model them as approaching flat surfaces, with the surrounding electrolyte ``sandwiched'' in-between. This description is particularly suitable in cases of short particle separations, whereby effects from surface curvature can always be neglected. Approximations that take curvature effects into account, such as the classical Derjaguin approximation might be also implemented once the interaction between the flat surfaces are known~\cite{Der34,Mc80,San19}. Even though both models are equally correct from a theoretical perspective, it is therefore of paramount importance to have a solid physical understanding on which one is more suitable for a particular application, as they can lead to quite different results. 
	
	Using a new MC technique, we have also investigated the phenomenon of local electroneutrality breakdown, and how it is influenced by the presence of patterned charge domains. It is shown that the presence of such surface charge modulations have a minor effect on the monopole charge distributions in the case of monovalent ions. The results indicate that these effects might be relevant in the presence of multivalent ions. This situation can not be captured by the employed linear approach. In fact, a deeper analysis of the effects of explicit osmotic or a Donnan-like equilibrium on the induced surface interactions would require the use of more sophisticated approaches able to incorporate non-linear and correlation effects, such as a Poisson-Boltzmann approach or a Density Functional Theory~\cite{Tam98}. It is also important to remark that the employed linear description incorporates neither ionic correlations nor non-linear interactions. While inclusion of such effects considerably increases the numerical complexity and rule out the advantage of working out explicit formulas for the induced interactions, we argue that these effects should not affect our main predictions, and should further extend the region in which electroneutrality takes place to larger inter-surface separations. Specifically, strong positional correlations tend to reduce the amount of counterions able to effectively pack together at the inner surface layers, thereby reducing their ability to screen the surface charges. A similar effect should be observed upon inclusion of solvent-ion size interactions at the confined electrolyte \cite{Jar09}. A detailed study of these effects on the surface interactions goes beyond the scope of this work, and might be the subject of future investigations. As a final remark, we point out that the Donnan potential can be explicitly computed by considering a Green Function method that explicitly incorporates different boundary conditions into the Poisson-Boltzmann equation. Work along this line is currently in progress. 
	
	\appendix
	
	\section{Osmotic stress between the charged surfaces}\label{Appendix1}
	
	We now provide a general demonstration of Eqs. (\ref{eq17}), (\ref{eq19}) and (\ref{eq20}), based on simple statistical mechanical arguments. To this end, we start with the definition of the osmotic stress across the plates, $\beta\Pi=-\dfrac{\partial\beta\mathcal{F}}{\partial d}$ (see Eq. (\ref{eq16})). The Free Energy is $\beta\mathcal{F}=-\log\Xi$, with $\Xi$ being the classical partition function,
	\begin{equation}
	\Xi=Tr[e^{-\beta \mathcal{H}}]
	\end{equation}
	where $Tr$ is the classical trace over ionic degrees of freedom. The potential energy comprises both internal ionic interactions $U_{in}$, and the external interactions $\Phi$ due to the presence of the charged walls. The internal contribution is
	\begin{equation}
	\beta U_{in}=\sum_{i<j}\beta u^{c}_{ij}({r}_ij)+\beta u^{hs}_{ij}({r}_ij),
	\end{equation}
	where $\beta u_{ij}^{c}(r_{ij})\equiv \frac{\lambda_B\alpha_1\alpha_j}{r_{ij}}$ and $u_{ij}^{hs}(r_{ij})\equiv 0$ if $r_{ij}\ge 2r$ and  $u_{ij}^{hs}(r_{ij})\equiv \infty$ if $r_{ij}< 2r$ denote the ionic Coulomb and hard-sphere pair interactions, respectively, of ions located at a separating distance $r_{ij}=|\bm{r}_i-\bm{r}_j|$. The external interactions, due to the presence of the surfaces, can be also split into hardcore and electrostatic contributions. The hardcore wall-ion interactions $\Phi^{hc}$ only depends on the transversal ion-membrane separations, and are given by
	\begin{eqnarray}
	\beta \Phi^{hc} & = & \sum_i\int\tilde{\rho}_i(\bm{S},z)\beta\phi_i^{hc}(z)d\bm{S}dz\label{pot_hc} \\
	\beta \Phi^{el} & = & \beta\Phi^{el}_1+\beta\Phi^{el}_2\nonumber\\
	& = & \int\phi^{el}_1(\bm{r})\left[\sum_i\alpha_i\tilde{\rho}_i(\bm{r})+\varrho_2(\bm{r})\right]d\bm{r}+\int\phi^{el}_2(\bm{r})\left[\sum_i\alpha_i\tilde{\rho}_i(\bm{r})+\varrho_2(\bm{r})\right]d\bm{r},
	\label{pot_el}
	\end{eqnarray}
	where $\tilde{\rho}_i(\bm{r})=\sum_{l\in i}\delta(\bm{r}-\bm{r}_l)$ is the static density of ionic component $i$. The hardcore wall-ion potential $\phi_i^{hc}(z)$ is:
	\begin{equation}
	\phi_i^{hc}(z)=
	\begin{cases}
	0, \qquad\text{if} \qquad m<|z|<l,\\
	\infty,\qquad\text{otherwise}.
	\end{cases}
	\label{phi_hc}
	\end{equation}
	where $m=d/2-s$ and $l=d/2+s$ are the inner/outer ion-surface closest separations. Denoting the charge density due to the fixed charged plates as $\varrho_{\substack{1\\2}}(\bm{r})=\sigma_{\substack{1\\2}}(\bm{S})\delta(z\pm d/2)$, the electrostatic potential produced by walls $1$ and $2$ can be formally written as:
	\begin{subequations}
		\begin{align}
		\beta\phi^{el}_1(\bm{r}) & =  \lambda_B\int\dfrac{\varrho_{1}(\bm{r}')}{|\bm{r}-\bm{r}'|}d\bm{r}'\label{pot_el1}\\
		\beta\phi^{el}_2(\bm{r}) & =  \lambda_B\int\dfrac{\varrho_{2}(\bm{r}')}{|\bm{r}-\bm{r}'|}d\bm{r}'\label{pot_el2}.
		\end{align}
	\end{subequations}
	
	The system Hamiltonian can be written as $\mathcal{H}=\mathcal{H}_0+\Phi^{el}+\Phi^{hc}$, where $\mathcal{H}_0$ is the Hamiltonian in the absence of the charged walls, comprising only ionic interactions and momenta. Notice that $\mathcal{H}_0$ is thus independent of the wall distance $d$. The osmotic pressure can thus be written as
	\begin{equation}
	\beta \Pi=-\dfrac{1}{A}\dfrac{\partial \beta\mathcal{F}}{\partial d}=\dfrac{1}{A\Xi}\dfrac{\partial}{\partial d}\left[Tr\left(e^{-\beta(\mathcal{H}_0+\Phi^{el}+\Phi^{hc})}\right)\right].
	\end{equation}
	Performing the differentiation leads to
	\begin{align}
	\beta \Pi=-\dfrac{1}{\Xi A}Tr\left[\left(\dfrac{\partial \beta\Phi^{hc}}{\partial d}+\dfrac{\partial \beta\Phi^{el}}{\partial d}\right) e^{-\beta(\mathcal{H}_0+\Phi^{el}+\Phi^{hc})}\right]=-\dfrac{1}{A}\left[\left\langle\dfrac{\partial \beta\Phi^{hc}}{\partial d}\right\rangle+\left\langle\dfrac{\partial \beta\Phi^{el}}{\partial d}\right\rangle\right],
	\end{align}
	where $\langle.\rangle$ denotes an ensemble average. Making now use of Eqs. (\ref{pot_hc}) and (\ref{pot_el}), the relation above becomes
	\begin{eqnarray}
	\beta \Pi  &\equiv & \beta\Pi^{hc}+\beta\Pi^{el},\label{Pi_1}\\
	\beta \Pi^{mec} & =  &-\dfrac{1}{A}\sum_i\int\dfrac{\partial \beta\phi_i^{hc}(z)}{\partial d}\rho_i(\bm{S},z)d\bm{S}dz,\label{Pi_hc1}\\
	\beta \Pi^{el} & = & -\dfrac{1}{A}\dfrac{\partial}{\partial d}\biggl[\int \left(\phi_1^{el}(\bm{r})+\phi^{el}_2(\bm{r})\right)\biggl(\sum_i\alpha_i{\rho}_i(\bm{r})\biggl)d\bm{r}+\int\phi_1^{el}(\bm{r})\varrho_2(\bm{r})d\bm{r}\biggl].\label{Pi_el1}
	\end{eqnarray}
	where $\varrho_{1}(\bm{r})=\sigma_1(\bm{S})\delta(z+d/2)$ and $\varrho_{2}(\bm{r})=\sigma_2(\bm{S})\delta(z-d/2)$, and $\rho_i(\bm{r})=\langle\tilde{\rho}_i(\bm{r})\rangle$ are the averaged ionic profiles. Notice that the surface electrostatic potentials, Eqs. (\ref{pot_el1}) and (\ref{pot_el2}), depends explicitly on the surface-surface distance $d$ only through the charge densities $\varrho_1(\bm{r})$ and $\varrho_2(\bm{r})$. The electrostatic interactions with the flat surfaces comprises both surface-ion and surface-surface interactions. Taking the derivative in Eq. (\ref{Pi_el1}) thus provides
	\begin{align}
	\beta\Pi^{el}=-\dfrac{1}{A}\int \biggl(\dfrac{\partial \beta\phi_1^{el}(\bm{r})}{\partial d}(\rho_\ell(\bm{r})-\varrho_1(\bm{r}))d\bm{r}
	+\dfrac{\partial \beta\phi_2^{el}(\bm{r})}{\partial d}(\rho_\ell(\bm{r})-\varrho_2(\bm{r}))\biggl)d\bm{r}
	\end{align} 
	where $\rho_\ell(\bm{r})\equiv \sum_i\alpha_i{\rho}_i(\bm{r})+\varrho_1(\bm{r})+\varrho_2(\bm{r})$ is the averaged net charge density at position $\bm{r}$. Making now usage of Eqs. (\ref{pot_el1}) and (\ref{pot_el2}), the above relation is simplified to:
	\begin{eqnarray}
	\beta\Pi^{el}=-\dfrac{1}{A}\biggl[\int \dfrac{\partial\varrho_1(\bm{r})}{\partial d}\left(\psi(\bm{r})-\phi_1^{el}(\bm{r})\right)d\bm{r}+\int \dfrac{\partial\varrho_2(\bm{r})}{\partial d}\left(\psi(\bm{r})-\phi_2^{el}(\bm{r})\right)d\bm{r}\biggl],
	\label{Pi_el2}
	\end{eqnarray} 
	where $\psi(\bm{r})$ is the mean electrostatic potential at position $\bm{r}$. Notice that the self-energy of the plates has been excluded from the surface interactions. The derivative of the surface charge densities can be explicitly evaluated:
	\begin{equation}
	\dfrac{\partial\varrho_{\substack{1\\2}}}{\partial d}=\sigma_{\substack{1\\2}}(\bm{S})\dfrac{\partial\delta(z\pm d/2)}{\partial d}=\pm\dfrac{1}{2}\sigma_{\substack{1\\2}}(\bm{S})\dfrac{\partial\delta(z\pm d/2)}{\partial z}.
	\end{equation}
	Substitution of the above result into Eq. (\ref{Pi_el2}) followed by an integration by parts along the $z$ coordinate leads to the following simplified relation for the electrostatic osmotic contribution to the inter-surface osmotic pressure:
	\begin{align}
	\beta\Pi^{el}=-\dfrac{1}{2A}\biggl[\int \sigma_1(\bm{S})\left(E_z(\bm{S})-E_z^{(1)}(\bm{S})\right)_{-d/2}d\bm{S}+\int \sigma_2(\bm{S})\left(E_z(\bm{S})-E_z^{(2)}(\bm{S})\right)_{d/2}d\bm{S}\biggl],
	\label{Pi_el2}
	\end{align} 
	where $E_z(\bm{S},z)=-\dfrac{\partial\psi}{\partial z}$ is the $z$ component of the total electrostatic field. Likewise, $E_z^{(1)}(\bm{S})=-\dfrac{\partial\phi^{el}_1}{\partial z}$ and $E_z^{(2)}(\bm{S})=-\dfrac{\partial\phi^{el}_2}{\partial z}$ are the electric fields produced by surfaces $1$ and $2$, respectively. Notice that these contributions should be calculated at the surface's contact. Even though the total field $E_z(\bm{S},z)$ is not defined in this point due to the surface charge discontinuity, the terms in brackets above remain continuous when we approach the surface from both sides. The above contribution to the osmotic pressure can also be written as
	\begin{align}
	\beta\Pi^{el}=-\dfrac{1}{2A}\biggl[\int \sigma_1(\bm{S})\left(\dfrac{E_z(\bm{S},-d/2^{+})+E_z(\bm{S},-d/2^{-})}{2}\right)d\bm{S}\nonumber\\+\int \sigma_2(\bm{S})\left(\dfrac{E_z(\bm{S},d/2^{+})+E_z(\bm{S},d/2^{-})}{2}\right)d\bm{S}\biggl],
	\label{Pi_el3}
	\end{align}
	where the terms in brackets are the averaged (total) electric fields across plates $1$ and $2$, respectively. The electric contribution to the osmotic pressure can be also written in Fourier space as
	
	\begin{align}
	\beta\Pi^{el}= \dfrac{(2\pi)^2}{2A}\biggl[\int\hat{\sigma}_2(\bm{q})\left(E_z(-\bm{q},d/2)-E_z^{(2)}(-\bm{q},d/2)\right)d\bm{q}\nonumber\\-\int\hat{\sigma}_1(\bm{q})\left(E_z(-\bm{q},-d/2)-E_z^{(1)}(-\bm{q},-d/2)\right)d\bm{q}\biggl].
	\label{pi1}
	\end{align}
	
	Now, the mechanical contribution $\beta\Pi^{mec}$ can be evaluated by first considering the derivative of the hardcore surface-ion pair potential. Using Eq. (\ref{phi_hc}), this derivative can be evaluated as
	\begin{align}
	&\dfrac{\partial\beta\phi^{hc}}{\partial d}  = -e^{\beta\phi^{hc}}\dfrac{\partial}{\partial d}(e^{-\beta\phi^{hc}})\\&=-\dfrac{\partial}{\partial d}\left[\Theta(-z-L)+\Theta(z+m)-\Theta(z-m)+\Theta(z-L)\right],
	\end{align}
	where $\Theta(x)$ denotes the usual Heaviside step-function. Performing the derivative, the above relation becomes
	\begin{equation}
	\dfrac{\partial\beta\phi^{hc}}{\partial d}=\dfrac{1}{2}\left[\delta(-z-L)-\delta(z+m)-\delta(z-m)+\delta(z-L)\right]. 
	\end{equation}
	Substitution into Eq. (\ref{Pi_hc1}) provides
	\begin{equation}
	\beta\Pi^{mec}=\dfrac{1}{2A}\int \left(\delta\rho_1(\bm{S})-\delta\rho_2(\bm{S})\right)d\bm{S},
	\end{equation}
	where $\delta\rho_1(\bm{S})\equiv\rho(\bm{S},z=-m)-\rho(\bm{S},z=-L)$ and $\delta\rho_2(\bm{S})\equiv\rho(\bm{S},z=L)-\rho(\bm{S},z=m)$ are the ionic density discontinuities across surfaces $1$ and $2$, respectively. In Fourier space, the expression above simplifies to
	\begin{equation}
	\beta\Pi^{hc}=\dfrac{(2\pi)^2}{2A}\left[\delta\hat{\rho}_1(\bm{q}=0)-\delta\hat{\rho}_2(\bm{q}=0)\right].
	\end{equation}
	In the context of a linearized DH theory, it follows that $\delta\hat{\rho}_1(\bm{q})=-\sum_i\rho^0_i\alpha_i\delta\psi_1(\bm{q})$ and $\delta\hat{\rho}_2(\bm{q})=-\sum_i\alpha_i\delta\psi_2(\bm{q})$, where $\delta\hat{\psi}_1(\bm{q})\equiv\hat{\psi}(\bm{q},-m)-\hat{\psi}(\bm{q},-L)$ and $\delta\hat{\psi}_2(\bm{q})\equiv\hat{\psi}(\bm{q},L)-\hat{\psi}(\bm{q},m)$. Due to the electroneutrality condition, $\sum_i\rho^0_i\alpha_i=0$, this hard-core contribution will always vanish in the DH level of approximation. Notice that the relations above for electrostatic and mechanical contributions to the osmotic stress have been obtained using an exact thermodynamic route, and should hold for any approach used for computing the ionic distributions.  Similar expressions for the case of homogeneous charge distributions have been also obtained using integral equations techniques~\cite{Loz84} and in the context of a density functional theory~\cite{colla}.

	\section{Explicit expressions for the averaged potentials}\label{Appendix2}

	We now provide the explicit solutions for the DH averaged potentials in the case of an explicit (Eq. (\ref{eq01})) and implicit (Eq. (\ref{eq33})).
	
	\subsection{Explicit reservoir model}
	
	We start by defining a parameter $\Delta$ as
	\begin{equation}
	\Delta\equiv(k+q)^2\sinh^2(2qs)f(2km)+2qk\left(f(2qs)-qke^{-2qs}\right)e^{2(km-qs)},
	\label{a21}
	\end{equation}
	where $l=d/2+s$, $m=d/2-s$ are the outer and inner ionic close approach positions ($s=z_m+r$ being the closest surface-ion distance), $f(x)$ and $g(x)$ are the functions defined in is defined in Eqs. (\ref{eq29}) and (\ref{eq39}), respectively.
	
	The potential below plate $1$ is
	
	\begin{equation}
	\hat{\psi}(\bm{q},z) = \dfrac{4\pi\lambda_Be^{k(z+l)}g(s)}{f(2qs)\Delta}\biggl[f(2qs)\hat{\sigma}_2(\bm{q})+(k^2-q^2)\sinh(2qs)e^{-2km}\hat{\sigma}_{1}(\bm{q})\biggl].
	\label{a22}
	\end{equation}  
	Across plate $1$, ($-l\le z\le -m$), it takes the form:
	
	\begin{align}
	&\hat{\psi}(\bm{q},z) =  \dfrac{4\pi\lambda_Bg(s)g(z+L)}{k\Delta f(2qs)}\biggl[(k^2-q^2)\sinh(2qs)e^{-2km}\hat{\sigma}_1(\bm{q})+f(2qs)\hat{\sigma}_2(\bm{q})\nonumber\\&+\dfrac{\hat{\sigma}_1\Delta}{kq}\biggl],\qquad(z\le -d/2)\label{a23}\\
	&\hat{\psi}(\bm{q},z) = \dfrac{4\pi\lambda_Bg(s)}{kf\Delta(2qs)}\biggl[\biggl((k^2-q^2)\sinh(2qs)e^{-2km}\hat{\sigma}_1(\bm{q})+f(2qs)\hat{\sigma}_2(\bm{q})\biggl) g(z+l)\nonumber\\&+\dfrac{\hat{\sigma}_1(\bm{q})\Delta}{kq}g(-z-m)\biggl], \qquad(z >-d/2).\label{a24}
	\end{align}
	In the inter-plate region, it is given by:
	\begin{align}
	&\hat{\psi}(\bm{q},z)=\dfrac{4\pi\lambda_Bg(s)}{\Delta k}\biggl[\hat{\sigma}_1(\bm{q})[g(2qs)e^{k(z-m)}-f(2qs)\sinh\left(k(z-m)\right)]\nonumber \\&+\hat{\sigma}_2(\bm{q})[g(2qs)e^{-k(z+m)}+f(2qs)\sinh(k(z+m))]\biggl].\label{a25}
	\end{align}
	Across plate $2$, ($m\le z\le l$), we have
	\begin{align}
	&\hat{\psi}(\bm{q},z) = \dfrac{4\pi\lambda_Bg(s)}{kf(2qs)\Delta}\biggl[\biggl(f(2qs)\hat{\sigma}_1(\bm{q})+(k^2-q^2)\sinh(2qs)e^{-2km}\hat{\sigma}_2(\bm{q})\biggl)g(L-z)\nonumber\\&+\dfrac{\hat{\sigma}_2(\bm{q})\Delta}{kq}g(z-m)\biggl],\qquad(z\le d/2)\label{a26}\\
	&\hat{\psi}(\bm{q},z)=  \dfrac{4\pi\lambda_Bg(s)g(L-z)}{kf(2qs)\Delta}\biggl[f(2qs)\hat{\sigma}_1(\bm{q})+(k^2-q^2)\sinh(2qs)e^{-2km}\hat{\sigma}_2(\bm{q})\nonumber\\&+\dfrac{\hat{\sigma}_2(\bm{q})\Delta}{kq}\biggl],\qquad(z > d/2).\label{a27}
	\end{align}
	Finally, in the region beyond plate $2$ ($z\ge l$), the potential takes the form:
	\begin{align}
	&\hat{\psi}(\bm{q},z)=\dfrac{4\pi\lambda_Bqe^{-k(z-l)}g(s)}{f(2qs)\Delta}\biggl[f(2qs)\hat{\sigma}_1(\bm{q})+(k^2-q^2)\sinh(2qs)e^{-2km}\hat{\sigma}_2(\bm{q})+\dfrac{\hat{\sigma}_2(\bm{q})\Delta}{kq}\biggl]\label{a28}
	\end{align}
	
	In the expressions above, $\hat{\sigma}_1$ and $\hat{\sigma}_2$ stands for the multipole surface charge densities $\hat{\sigma}_h(\bm{q})$ at plates $1$ and $2$, respectively. Similar expressions can be obtained for the monopole averaged potential by taking the limit $q\rightarrow 0$ in the above expressions.
	
	\subsection{Explicit reservoir model}
	
	The potential below the first surface ($z\le -d/2$) is
	\begin{align}
	&\hat{\psi}(\bm{q},z)=\dfrac{4\pi\lambda_B}{qf(2ks)} e^{q(z+m)}\biggl[\hat{\sigma}_1(\bm{q})\left(f(2km)\cosh(qs)-g(2km)e^{-qs}\right)+kqe^{-qs}\hat{\sigma}_2(\bm{q})\biggl].
	\label{a30}
	\end{align}
	Across plate $1$, ($-d/2\le z <-m$) the potential is given by
	\begin{align}
	&\hat{\psi}(\bm{q},z)=\dfrac{4\pi\lambda_B e^{-qs}}{qf(2km)}\biggl[\biggl(f(2km)\cosh[q(z+m)]-g(2km)e^{q(z+m)}\biggl)\hat{\sigma}_1(\bm{q})+kq\hat{\sigma}_2(\bm{q})e^{q(z+m)}\biggl],
	\label{a31}
	\end{align}
	while at the inter-surface space ($-m\le z < m$) it takes the form:
	\begin{align}
	&\hat{\psi}(\bm{q},z)=\dfrac{4\pi\lambda_Be^{-qs}}{f(2km)}\biggl[\biggl(k\cosh(k(z-m))-q\sinh(k(z-m))\biggl)\hat{\sigma}_1(\bm{q})+\nonumber\\&\biggl(k\cosh(k(z+m))+q\sinh(k(z+m))\biggl)\hat{\sigma}_2(\bm{q})\biggl].
	\label{a32}
	\end{align}
	Across the second plate ($m\le z < d/2$), we have:
	\begin{align}
	&\hat{\psi}(\bm{q},z)= \dfrac{4\pi \lambda_Be^{-qs}}{qf(2km)}\biggl[kq\hat{\sigma}_1(\bm{q})e^{-q(z-m)}+\biggl(f(2km)\cosh(q(z-m))-g(2km)e^{-q(z-m)}\biggl)\hat{\sigma}_2(\bm{q})\biggl].
	\label{a33}
	\end{align}
	Finally, the region beyond plate $2$ ($z\ge d/2$) has the following electrostatic potentials:
	\begin{align}
	&\hat{\psi}(\bm{q},z)=\dfrac{4\pi\lambda_B}{qf(2km)} e^{-q(z-m)}\biggl[\hat{\sigma}_2(\bm{q})\left(f(2km)\cosh(qs)-g(2km)e^{-qs}\right)+kqe^{-qs}\hat{\sigma}_1(\bm{q})\biggl].
	\end{align}
	Explicit relations for the monopole contributions can be readily obtained by taking the limit $q\rightarrow 0$ in expressions (\ref{a31}), (\ref{a32}) and (\ref{a33}).

	\section*{References}
	%\bibliographystyle{elsarticle-num-names}
	%\bibliography{ref.bib}
	%\begin{thebibliography}{68}%
	
	%\end{thebibliography}%

	%

\end{document}